\documentclass[usenatbib,letterpaper,hyperref]{mn2e}
\usepackage[totalwidth=480pt, totalheight=680pt]{geometry}
\usepackage{hyperref}
\usepackage{times}
\usepackage{epsfig}
\usepackage{amsmath}
\usepackage{amssymb}
\usepackage{url}
\usepackage{aas_macros}
\def\eqsim{\mathrel{\raise0.35ex\hbox{$\scriptstyle =$}\kern-0.6em
    \lower0.40ex\hbox{{$\scriptstyle \sim$}}}}
\def\gtrsim{\mathrel{\raise0.35ex\hbox{$\scriptstyle >$}\kern-0.6em
    \lower0.40ex\hbox{{$\scriptstyle \sim$}}}}
\def\lesssim{\mathrel{\raise0.35ex\hbox{$\scriptstyle <$}\kern-0.6em
    \lower0.40ex\hbox{{$\scriptstyle \sim$}}}}

\title[The low-acceleration RAR]{Observational constraints on the slope of the radial acceleration relation at low accelerations}
\author[K. A. Oman et al.]{
  \newauthor Kyle A. Oman$^{1}$\thanks{kyle.a.oman@durham.ac.uk}, Margot M. Brouwer$^{2,3}$, Aaron D. Ludlow$^{4}$ and Julio F. Navarro$^{5}$\\
  $^{1}$Institute for Computational Cosmology, Department of Physics, Durham University, South Road, Durham DH1 3LE, United Kingdom\\
  $^{2}$Kapteyn Astronomical Institute, University of Groningen, Postbus 800, NL-9700 AV Groningen, The Netherlands\\
  $^{3}$Institute for Theoretical Physics, University of Amsterdam, Science Park 904, 1098 XH Amsterdam, The Netherlands\\
  $^{4}$International Centre for Radio Astronomy Research, University of Western Australia, 35 Stirling Highway, Crawley, Western Australia 6009,\\
  Australia\\
  $^{5}$Department of Physics and Astronomy, University of Victoria, Victoria, BC V8P 5C2, Canada
}
\date{\today}

\begin{document}
\label{firstpage}
\maketitle

\begin{abstract} 
  The radial acceleration relation (RAR) locally relates the `observed' acceleration inferred from the dynamics of a system to the acceleration implied by its baryonic matter distribution. The relation as traced by galaxy rotation curves is one-to-one with remarkably little scatter, implying that the dynamics of a system can be predicted simply by measuring its density profile as traced by e.g. stellar light or gas emission lines. Extending the relation to accelerations below those usually probed by practically observable kinematic tracers is challenging, especially once accounting for faintly emitting baryons, such as the putative warm-hot intergalactic medium, becomes important. We show that in the low-acceleration regime, the (inverted) RAR predicts an unphysical, declining enclosed baryonic mass profile for systems with `observed' acceleration profiles steeper than $g_{\rm obs}\propto r^{-1}$ (corresponding to density profiles steeper than isothermal -- $\rho(r)\propto r^{-2}$). If the RAR is tantamount to a natural law, such acceleration profiles cannot exist. We apply this argument to test the compatibility of an extrapolation of the rotation curve-derived RAR to low accelerations with data from galaxy-galaxy weak lensing, dwarf spheroidal galaxy stellar kinematic, and outer Milky~Way dynamical measurements, fully independent of the uncertainties inherent in direct measurements of the baryonic matter distribution. In all cases we find that the data weakly favour a break to a steeper low-acceleration slope. Improvements in measurements and modelling of the outer Milky~Way, and weak lensing, seem like the most promising path toward stronger constraints on the low-acceleration behaviour of the RAR.
\end{abstract}

\section{Introduction}
\label{SecIntro}

\begin{figure}
  \includegraphics[width=\columnwidth]{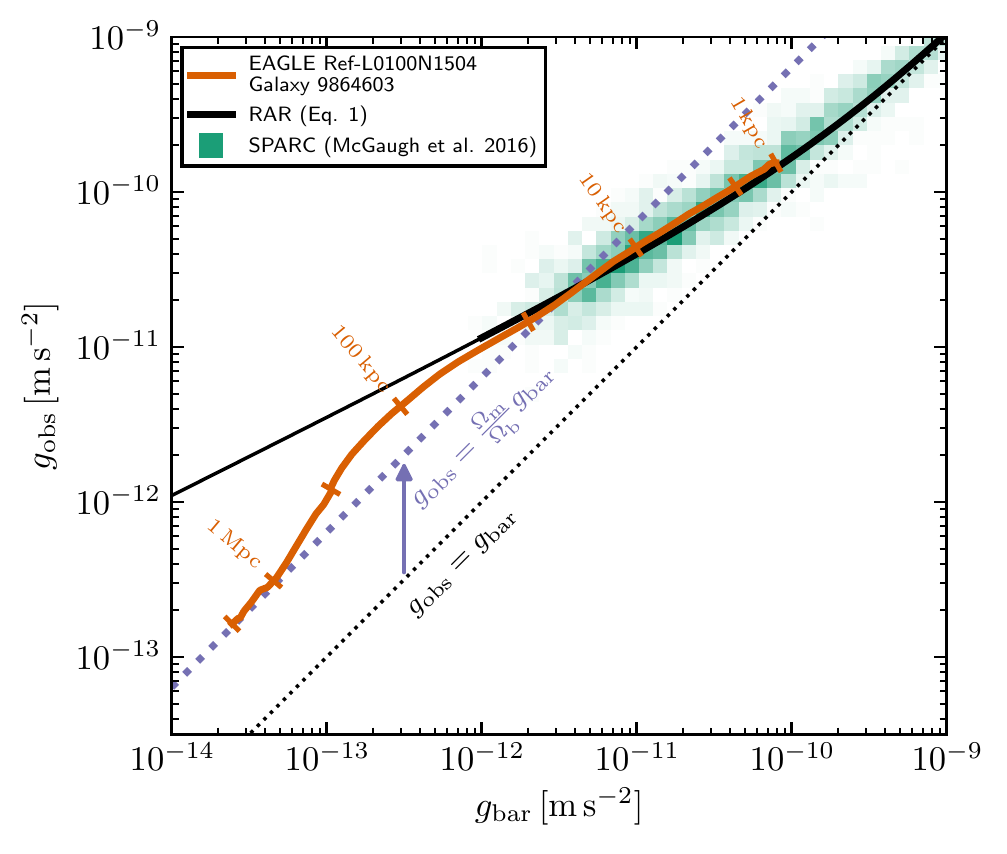}
  \caption{Illustrative example of the expectation for the RAR in different theoretical frameworks. The data and proposed fitting function (Eq.~\ref{eq-rar}) of \citet[][fig.~3]{2016PhRvL.117t1101M} are reproduced with the heatmap and solid black line, respectively; the fitting function is continued outside the range spanned by the data with a thinner line. This curve represents the simplest expectation in some modified gravity theories such as MOND and EG. To illustrate the qualitative expectation in a $\Lambda{\rm CDM}$ cosmology, we use the galaxy with identifier $9864603$ (orange solid line) from the EAGLE Ref-L0100N1504 cosmological hydrodynamical simulation \citep{2015MNRAS.446..521S,2009MNRAS.399.1773C}. We calculate the accelerations as $g_{\rm obs}=\frac{GM(<r)}{r^2}$ and $g_{\rm bar}=\frac{GM_{\rm bar}(<r)}{r^2}$. The labelled ticks at $0.5\,{\rm dex}$ intervals mark the radius in the example galaxy. At low accelerations ($g_{\rm obs}\lesssim 10^{-11}\,{\rm m}\,{\rm s}^{-2}$), the curve for the simulated galaxy departs from the form of Eq.~\ref{eq-rar} and converges to the heavy dotted line, corresponding to the expectation for a cosmological baryon-to-total matter ratio $f_b=\frac{\Omega_{\rm b}}{\Omega_{\rm m}}$. EAGLE galaxies are, on average, systematically slightly offset from the RAR \citep[roughly speaking, $g_\dagger$ differs slightly from the observed value, see][for details]{2017PhRvL.118p1103L}; the galaxy shown here has been chosen as an example that closely follows the RAR at accelerations $g_{\rm bar}\gtrsim 10^{-12}\,{\rm m}\,{\rm s}^{-2}$.}
  \label{fig-eagle-rar}
\end{figure}

\citet{2016PhRvL.117t1101M} used galaxy rotation curve measurements to show that there is a tight relationship between the `observed' local circular acceleration, $g_{\rm obs}$, within a galaxy, and the acceleration expected at the same location given the visible distribution of matter, $g_{\rm bar}$. They show (reproduced in Fig.~\ref{fig-eagle-rar}) that the following functional form provides a good fit to the data for late-type galaxies over the observed range in accelerations ($10^{-12}\lesssim g_{\rm bar}/{\rm m}\,{\rm s}^{-2}\lesssim 10^{-8}$):
\begin{align}
g_{\rm obs} = \frac{g_{\rm bar}}{1-e^{-\sqrt{g_{\rm bar}/g_\dag}}}\label{eq-rar}
\end{align}
with $g_{\dag}=1.2\times10^{-10}\,{\rm m}\,{\rm s}^{-2}$, which we will assume throughout this work\footnote{We will sometimes refer to Eq.~\ref{eq-rar} with this value of $g_\dag$ as the `fiducial RAR' below.}. They also note that the observed scatter in the relation seems to be consistent with originating entirely from measurement uncertainties, i.e. the intrinsic relation has nearly zero scatter, and \citet{2017ApJ...836..152L} further point out that the residuals from their fiducial relation do not correlate with stellar surface brightness, galaxy size, or gas fraction. At the high-acceleration end, $g_{\rm obs}\approx g_{\rm bar}$, that is, the observed acceleration is fully explained by the observed mass distribution. At the low-acceleration end, however, the observed acceleration is higher than the Newtonian expectation if light follows mass. This can be explained either by the presence of additional mass -- either dark matter or, more prosaically, baryons not yet accounted for -- or by a departure from the Newtonian theory of gravity in the low-acceleration limit.

The low-acceleration limit of Eq.~\ref{eq-rar} happens to be exactly that proposed in the theory of MOdified Newtonian Dynamics \citep[MOND;][]{1983ApJ...270..365M}, leading the RAR to be interpreted by some as strong evidence in support of MOND \citep[e.g.][]{2012LRR....15...10F,2012PASA...29..395K,2015CaJPh..93..250M}. The emergent gravity (EG) theory of \citet{2017ScPP....2...16V} also has the same\footnote{The transition between the high- and low-acceleration limits has a slightly different form than that of Eq.~\ref{eq-rar}, see Appendix~\ref{app-derive}} low-acceleration limit \citep{2017MNRAS.466.2547B}, and the Weyl conformal gravity theory also has a similar low-acceleration behaviour \citep{2020PhRvD.101h4015I}. Others argue that the RAR can be explained in $\Lambda{\rm CDM}$-based models of galaxy formation without fine-tuning beyond requiring that galaxies follow the observed scaling relations in mass, characteristic velocity, and size \citep[e.g.][]{2000ApJ...534..146V,2002ApJ...569L..19K,2016MNRAS.455..476S,2017ApJ...835L..17K,2017MNRAS.471.1841N,2017PhRvL.118p1103L,2019MNRAS.485.1886D,2019ApJ...882...46W}.

Since the initial definition of the RAR by \citet{2016PhRvL.117t1101M}, the claim that the acceleration scale $g_\dag$ in Eq.~\ref{eq-rar} is universal has been called into question \citep[e.g.][]{2018NatAs...2..668R,2018MNRAS.477..230R,2019ApJ...882....6S,2020MNRAS.492.5865C,2020arXiv200308845Z}. \citet{2019ApJ...882....6S} further suggest that the scatter may be much larger than claimed by \citet{2016PhRvL.117t1101M}. Though these observations may yield useful insight into the physical interpretation of the RAR, for the present study we will assume a fixed value of $g_{\dag}$, and negligible scatter.

Although the rotation curve-based measurement of the RAR \citep{2016PhRvL.117t1101M} extends $\sim 2$ decades in acceleration below $g_\dag$, there is a case to push to lower accelerations in order to discriminate between modified gravity and dark matter models. In MOND, for instance, the $g_{\rm obs}\propto \sqrt{g_{\rm bar}}$ scaling should continue to arbitrarily small accelerations, whereas in a universe with a baryon fraction $f_{\rm b}=\Omega_{\rm b}/\Omega_{\rm m} < 1$ the RAR should eventually bend again and converge to $g_{\rm obs}=g_{\rm bar}/f_{\rm b}$ on scales large enough that the enclosed ratio of baryonic and dark matter reliably converges to the cosmic mean, as illustrated in Fig.~\ref{fig-eagle-rar}. An extension of the RAR to sufficiently low accelerations should therefore be able to discriminate between theories. In practice both of these predictions are more complicated: in the MOND case, once acceleration fields external to the system under study become important, corrections for the external field must be taken into account \citep[][and references therein]{2000ApJ...541..556B}, while in $\Lambda{\rm CDM}$ neighbouring haloes constitute departures from spherical symmetry which must in principle be taken into account on the scales in question. Nevertheless, it seems clear that the low-acceleration behavior of the RAR should be qualitatively different in these two scenarios.

The RAR can be extended $\sim 2$~decades toward lower accelerations by including dwarf spheroidal (dSph) galaxies. \citet{2017ApJ...836..152L} attempted this and found tentative evidence for an upturn in the relation, but were hampered by concerns regarding the possible modified gravity corrections needed to account for the external acceleration field due to either the Milky~Way or Andromeda. It is challenging to extend the relation any further since measurements must be made either in extremely low-mass, intrinsically faint systems which are currently only observable in close proximity to more massive hosts, or at large distances around more massive systems. In the latter case, eventually the unknown distribution of baryons in the `warm-hot intergalactic medium', and other difficult-to-observe phases, becomes a dominant systematic uncertainty on $g_{\rm bar}$ \citep{1998ApJ...503..518F,2004ApJ...616..643F,2019A&A...624A..48D,2019MNRAS.487.2148G}: any upturn in the RAR at the low-acceleration end is suspect since $g_{\rm bar}$ is always implicitly a lower limit.

In Sec.~\ref{sec-arg} we outline a simple argument which shows that, if Eq.~\ref{eq-rar} holds, some acceleration profiles $g_{\rm obs}(r)$ have no corresponding physically allowed baryonic acceleration profiles $g_{\rm bar}(r)$, implying that a lower limit can be placed on the low-acceleration slope of the RAR \emph{without measuring $g_{\rm bar}$}. In Secs.~\ref{subsec-rc}--\ref{subsec-mw} we outline how this argument can be applied in practice for the cases of galaxy rotation curves, galaxy-galaxy weak gravitational lensing, stellar kinematics in dSph galaxies, and the orbits of satellites of the Milky~Way. We summarize in Sec.~\ref{sec-conc}.

\section{The RAR as a predictor of baryons}\label{sec-arg}

As it is usually presented (Eq.~\ref{eq-rar}), the RAR implies that the dynamics of a system can be predicted from its baryonic mass distribution: $g_{\rm obs}(g_{\rm bar})$. However, this function can be inverted in order to predict the baryonic mass distribution. The functional form proposed by \citet{2016PhRvL.117t1101M} is not analytically invertible, but is straightforward to invert numerically.

\begin{figure}
  \includegraphics[width=\columnwidth]{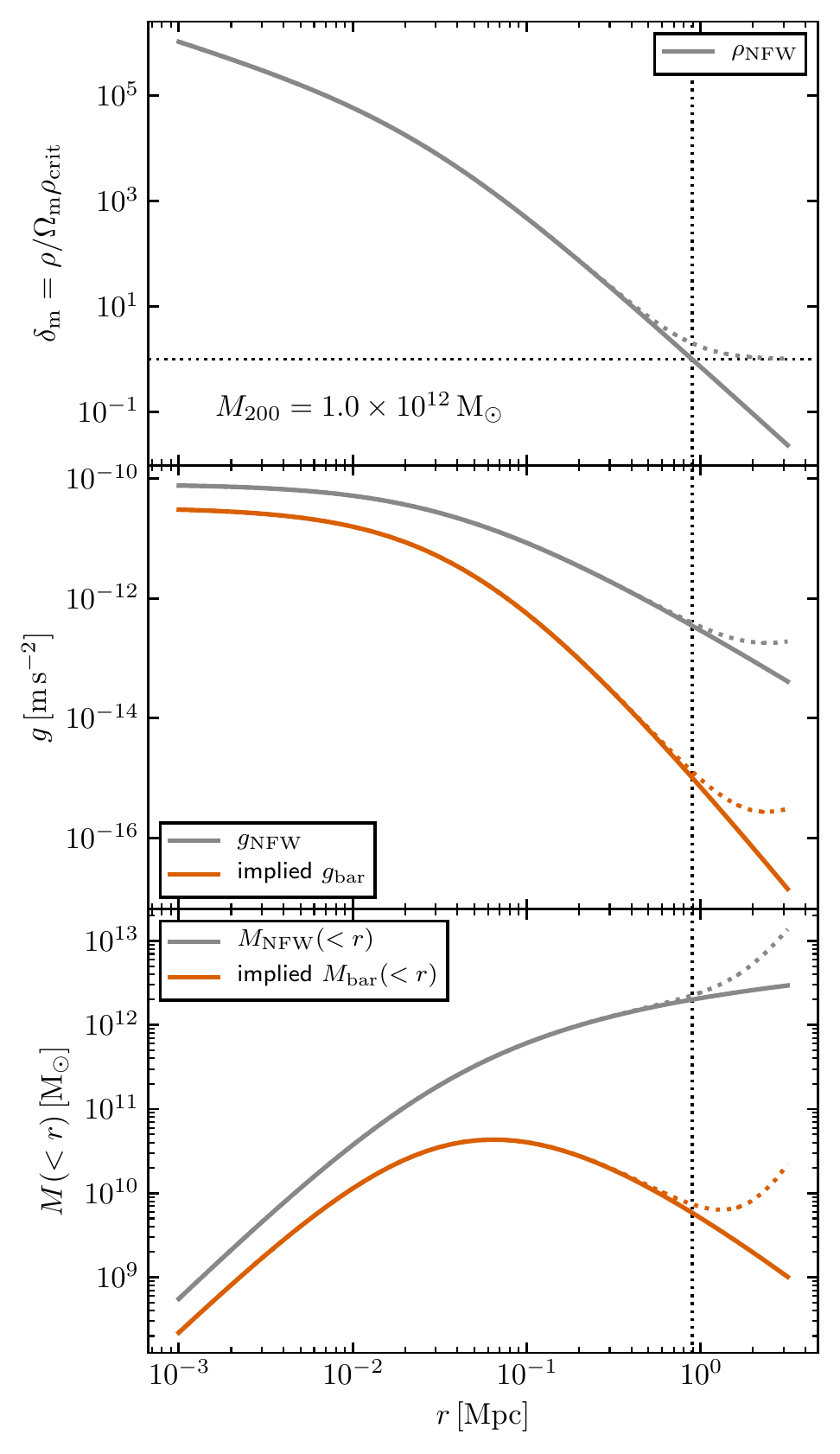}
  \caption{\emph{Upper: }Density profile of an NFW halo with $M_{200}=10^{12}\,{\rm M}_\odot$, in units of the mean matter density (grey line). The vertical dotted line marks where the profile crosses $\delta_{\rm m}=1$, as a crude estimate of the extreme outer edge of the halo model. The dotted grey line is the same profile, added to a constant background density equal to the cosmic mean. \emph{Centre: }Acceleration profile corresponding to the density profile in the upper panel (grey line). We solve Eq.~\ref{eq-rar} for $g_{\rm bar}$ assuming $g_{\rm obs}=g_{\rm NFW}$; the resulting $g_{\rm bar}$ is shown with the orange line. Dotted lines correspond to the profile including a constant background density. \emph{Lower: }Cumulative mass profiles corresponding to the acceleration profiles in the middle panel. For a $\Lambda{\rm CDM}$-motivated NFW acceleration profile, the RAR implies a physically nonsensical \emph{declining} cumulative mass of baryons at large radii (orange line).}
  \label{fig-nfw}
\end{figure}

In Fig.~\ref{fig-nfw}, we show an illustrative application of this inverted function. For simplicity, we assume spherical symmetry throughout this example. We use an NFW halo model \citep{1996ApJ...462..563N} as an example as it aptly illustrates the mutual incompatibility between the modified gravity and dark matter-based interpretations of the RAR. The upper panel shows the density profile such a model with a mass\footnote{The mass $M_{200}$ is defined as that enclosed within a sphere with a mean enclosed density $200\times$ the critical density for closure, $\rho_{\rm crit}=\frac{3H^2}{8\pi G}$.} $M_{200}=10^{12}\,{\rm M}_\odot$ and concentration $c=8$, in units of the mean matter density. The concentration is chosen such that the halo lies approximately on the mass-concentration relation of \citet{2014MNRAS.441..378L} for $z=0$. The vertical dotted line marks the radius where the halo density crosses the mean matter density\footnote{We adopt $\Omega_m=0.3$ and $H_0=70\,{\rm km}\,{\rm s}^{-1}\,{\rm Mpc}^{-1}$.}, intended as a crude estimate of the location of the extreme outer edge of the halo. The dotted line corresponds to the same NFW profile summed with a constant background density equal to the cosmic mean. The centre panel of Fig.~\ref{fig-nfw} shows with a grey line the corresponding acceleration profile:
\begin{align}
  g_{\rm NFW}(r)&=\frac{4\pi G}{r^2}\int_0^r\rho_{\rm NFW}(r')r'^2\,{\rm d}r'\\
  g_{\rm NFW}(r)&=\frac{GM_{\rm NFW}(<r)}{r^2}
\end{align}
where $M_{\rm NFW}(<r)$ is the mass enclosed within radius $r$ (shown with a grey line in the lower panel). We then solve Eq.~\ref{eq-rar} for the baryonic acceleration profile implied assuming\footnote{In a dark matter model it would seem natural to instead assume $g_{\rm obs}=g_{\rm NFW}+g_{\rm bar}$, which would lead to a qualitatively similar result below. However in alternative theories of gravity such as MOND, accelerations do not add linearly, so we assume $g_{\rm obs}=g_{\rm NFW}$ to preserve the generality of our example.} $g_{\rm obs}=g_{\rm NFW}$. The implied $g_{\rm bar}$, assuming only that Eq.~\ref{eq-rar} holds, is shown with the orange line. The dotted lines of corresponding colour include the constant background density, as in the upper panel. The corresponding cumulative mass profiles $M(<r)=g(r)r/G$ are shown in the lower panel. The implied $M_{\rm bar}(<r)$ (orange line) is clearly unphysical: a cumulative mass profile must be monotonically increasing.

The above example illustrates that some acceleration profiles, such as the $g_{\rm obs}(r)$ in Fig.~\ref{fig-nfw}, cannot arise in a universe where the RAR is taken to be a natural law. The conditions required to produce declining $M(r)$ profiles (which are unphysical) depend on the acceleration scale. When $g_{\rm bar}\gg g_\dag$, then $g_{\rm obs}\propto g_{\rm bar}$ and any physically reasonable observed acceleration profile should imply a physically reasonable $M_{\rm bar}(<r)$. In the low acceleration limit $g_{\rm bar}\ll g_\dag$, however, $g_{\rm obs}\propto\sqrt{g_{\rm bar}}$, which gives:
\begin{align}
  M_{\rm bar}(<r)\propto\left(\frac{M_{\rm obs}(<r)}{r}\right)^2.
\end{align}
While the above example is loosely inspired by a galaxy model including baryons and (non-baryonic) dark matter, this is not required for our argument to be valid. Clearly, any $M_{\rm obs}(<r)$ inferred to increase less steeply than linearly with radius will lead to a declining cumulative baryonic mass profile -- or, in terms of density, any profile inferred to fall off more steeply than isothermal ($\propto r^{-2}$) will be problematic.

One interesting implication of the above is that it is possible to test for some types of departures from Eq.~\ref{eq-rar} \emph{without measuring the baryonic acceleration profile}. Provided the system being considered is close to spherically symmetric and close to dynamical equilibrium, if its observed acceleration profile (i.e. inferred from a dynamical measurement) implies a density profile steeper than isothermal in a region where $g_{\rm obs}\ll g_\dag$, then\footnote{Since on the RAR $g_{\rm bar}\leq g_{\rm obs}$, $g_{\rm obs}\ll g_\dag$ implies $g_{\rm bar}\ll g_\dag$.} there is no physical baryonic mass profile that will place the system on the fiducial RAR.

In the next sections we explore possible observational signatures of acceleration profiles that would imply a departure from $g_{\rm obs}\propto\sqrt{g_{\rm bar}}$ in the low-acceleration limit\footnote{We give a derivation of the acceleration-dependent limit on the slope in Appendix~\ref{app-derive}.}.

\subsection{Rotation curves}\label{subsec-rc}

Applying the above argument to the speed $V_{\rm circ}(r)=\sqrt{GM_{\rm obs}(<r)/r}$ of a particle on a circular orbit in a spherically symmetric potential leads to a simple constraint on rotation curves compatible with the low-acceleration limit of the RAR. The implied baryonic mass profile is:
\begin{align}
  M_{\rm bar}(<r)&\propto (V_{\rm circ}(r))^4 \label{eq-btfr}.
\end{align}
Therefore any rotation curve $V_{\rm circ}(r)$ which is declining (${\rm d}V_{\rm circ}/{\rm d}r<0$) in a region where $g_{\rm obs}\ll g_\dag$ leads to a declining $M_{\rm bar}(<r)$. Note that regardless of whether Eq.~\ref{eq-rar} holds, rotation curves with slopes falling more steeply than ${\rm d}\log V_{\rm circ}/{\rm d}\log r<-\frac{1}{2}$ cannot be interpreted as tracers of the total mass profile, independent of the acceleration scale -- this limit corresponds to the Keplerian (point mass) circular velocity curve.

\begin{figure}
  \includegraphics[width=\columnwidth]{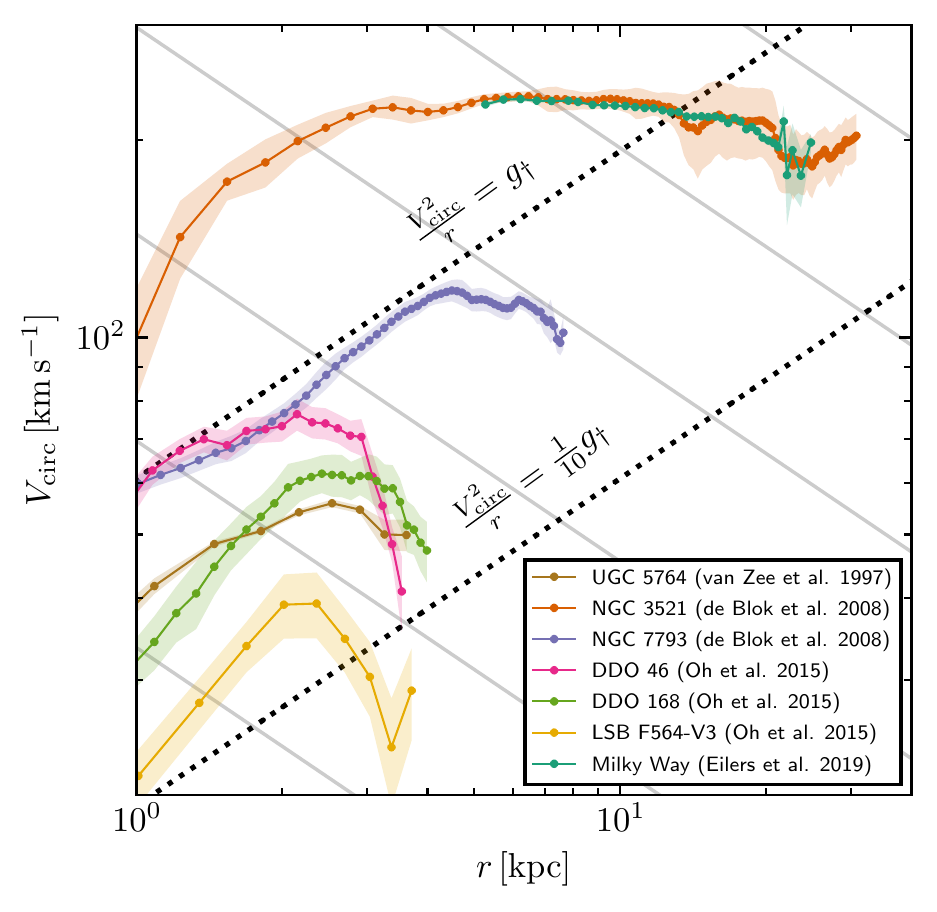}
  \caption{Selected literature rotation curves \citep{1997AJ....113.1618V,2008AJ....136.2648D,2015AJ....149..180O,2019ApJ...871..120E} with an apparent decline at low accelerations (none of which, with the possible exception of the Milky~Way, constitute convincing evidence for a departure from the fiducial RAR, see Sec.~\ref{subsec-rc}). The shaded regions correspond to the uncertainty intervals quoted in each source publication and are heterogenously defined. The dotted lines are curves of constant acceleration. The grey lines illustrate a slope ${\rm d}\log V_{\rm circ}/{\rm d}\log r=-\frac{1}{2}$ (Keplerian fall-off), the steepest decline possible for a circular velocity curve.}
  \label{fig-rc}
\end{figure}

Given that the RAR was defined by \citet{2016PhRvL.117t1101M} using galaxy rotation curves, we do not expect these to deviate systematically from the RAR. However, since even a single compelling example of a low-acceleration, declining rotation curve has implications for the physical interpretation of the RAR, we have searched archival rotation curves for any examples meriting further consideration. We have not identified any compelling measurements of low-acceleration, declining rotation curves \citep[see also][]{2008AJ....136.2648D,2016A&A...585A..99M}, but touch on a few debated examples for completeness. We show a small selection of apparently declining rotation curves in Fig.~\ref{fig-rc}, noting that the error bars shown correspond to the uncertainties quoted in the cited works and are not homogeneously defined.

\subsubsection{External galaxies}

The SPARC \citep{2016AJ....152..157L} compilation used by \citet{2016PhRvL.117t1101M} includes only one example of a potentially declining rotation curve at low accelerations, UGC~5764, drawn from \citet{1997AJ....113.1618V}, which we reproduce in Fig.~\ref{fig-rc}. However, the decline is only seen on one side of the disc and the reported rotation curve is consistent with a flat slope, within the reported uncertainties.

The study of ${\rm H}$\,{\sc i} kinematics of THINGS survey \citep{2008AJ....136.2563W} galaxies by \citet{2008AJ....136.2648D} discusses at length the possible examples of declining rotation curves in their sample, in the context of searching for the signature of the `edge' of dark matter haloes. We reproduce two examples from their analysis, NGC~3521 and NGC~7793, in Fig.~\ref{fig-rc}, which had previously been held up as rare examples of `truly declining rotation curves' \citep{1990AJ....100..394C,1991AJ....101.1231C}. However, \citet{2008AJ....136.2648D} state that an outer decline cannot be unambiguously established for any of their galaxies within the uncertainties on their rotation speeds and inclinations. For NGC~7793, they cite difficulty in modelling an apparent outer warp, while for NGC~3521 the combination of only a mild decline and relatively large uncertainties make the flat rotation curve hypothesis difficult to reject.

We also reproduce in Fig.~\ref{fig-rc} three rotation curves from the \citet{2015AJ....149..180O} analysis of the LITTLE THINGS survey \citep{2012AJ....144..134H}: those of DDO~46, DDO~168 and LSB~F564-V3. \citet{2015AJ....149..180O} did not comment on individual systems, however \citet{2017MNRAS.466.4159I} re-analyzed most of the same galaxies with a qualitatively different 3D fitting technique and commented on any differences in the results. They did not find a declining rotation curve for DDO~168, and pointed out several kinematic irregularities -- a clear bar in both the stellar and gas distributions, a `very irregular' outer disc, possibly warped, and possible signs of significant amounts of inflowing/outflowing gas -- which may account for the differences. DDO~168 also has a companion, DDO~167, at a projected separation of only 33~kpc \citep{2015AJ....149..196J}. Unfortunately neither DDO~46 nor LSB~F564-V3 appear in the \citet{2017MNRAS.466.4159I} analysis: the former was excluded due to its low inclination (and thus large and uncertain correction to the rotation velocity), while the reason for the exclusion of the latter is unclear. A rotation curve for LSB~F564-V3 was also determined by \citet[][note that in that work the name LSB~D564-8 is used]{2009A&A...505..577T}, who found no decline in the outer parts. Furthermore, inspection of figs.~A.49 \& A.7 in \citet{2015AJ....149..180O} reveals that the declining part of the rotation curves of DDO~46 and LSB~D564-8 occurs in regions of very low signal-to-noise ratio.

Finally, we note that the rotation curves of DDO~46, DDO~168 and LSB~F564-V3 cannot be interpreted as tracers of the enclosed mass, as they fall more steeply than the Keplerian limit (illustrated by the slope of the thin grey lines in Fig.~\ref{fig-rc}): if taken at face value, the declining rotation velocities in the outer parts of these galaxies imply a (nonsensical) declining cumulative dynamical mass profile.

We conclude that evidence in the literature for low-acceleration, declining rotation curves for external galaxies is marginal, at best. The unique case of the Milky~Way, however, may be more constraining, as we elaborate below.

\subsubsection{The Milky~Way}\label{subsubsec-mwrc}

We also reproduce the Milky~Way rotation curve of \citet{2019ApJ...871..120E} in Fig.~\ref{fig-rc}. Their rotation curve is based on 6-dimensional kinematic information for red giant stars, relying particularly on the APOGEE and Gaia surveys for spectroscopic and astrometric measurements, respectively. We discuss analyses of other kinematic tracers in the outer Milky~Way, outside the stellar disc, in Sec.~\ref{subsec-mw}.

\citet{2019ApJ...871..120E} reported a significant detection of a $-1.7\pm 0.1\,({\rm stat.})\pm 0.47\,({\rm sys.})\,{\rm km}\,{\rm s}^{-1}\,{\rm kpc}^{-1}$ gradient in rotation velocity out to a distance of $\sim 25\,{\rm kpc}$ \citep[see also][who find the same gradient using Cepheids, albeit with more local fluctuations along the rotation curve]{2020arXiv200413768A}. The region of the Milky~Way spanned by the \citet{2019ApJ...871..120E} analysis is in the transition region between the high- and low-acceleration limits of Eq.~\ref{eq-rar} (see Fig.~\ref{fig-rc}: the Milky~Way curve crosses the diagonal dashed line corresponding to $g_{\dag}$, marking the approximate location of the transition between the high- and low-acceleration regimes), so the limiting slope of the rotation curve -- or equivalently the acceleration profile $g_{\rm obs}(r)$ -- compatible with the fiducial RAR is varying with radius. We illustrate this in Fig.~\ref{fig-eilers}; further details are given in Appendix~\ref{app-derive}. In the upper panel, we show the acceleration profile of the Milky~Way, derived from the rotation curve as $g_{\rm obs}(r)=V_{\rm circ}(r)^2/r$, with points and error bars. The steepest ${\rm d}\log g_{\rm obs}/{\rm d}\log r$ compatible with Eq.~\ref{eq-rar} is a function of acceleration, we show the limiting slope with black line segments for two example points marked with black squares. At the higher acceleration example point, the Milky~Way acceleration profile is shallower than the limit, as illustrated by the orange line segment, which is tangent to the data at the same point. At the lower acceleration point, however, the orange tangent is steeper than the limit implied by Eq.~\ref{eq-rar}, which constitutes evidence for a downward bend away from the fiducial relation.

We further quantify this evidence in the lower panel of Fig.~\ref{fig-eilers}. The slope of the fiducial RAR (Eq.~\ref{eq-rar}) as a function of acceleration is shown with the black curve. As the argument laid out in Sec.~\ref{sec-arg} yields a lower limit on ${\rm d}\log g_{\rm obs}/{\rm d}\log g_{\rm bar}$, a constraint lying above this curve -- in the grey shaded region -- implies a break to a steeper low-acceleration slope than that of Eq.~\ref{eq-rar}. To derive the constraint on the slope as a function of acceleration $g_{\rm obs}/g_\dag$, we assume the linear fit to the rotation curve given in eq.~7 of \citet{2019ApJ...871..120E}, with the solar radius $R_\odot=8.122\,{\rm kpc}$. We neglect the uncertainties in the solar radius and normalization of the linear fit as the constraint on the RAR slope is minimally sensitive to these. For simplicity, we assume that the statistical and systematic uncertainties on the slope measurement both follow a Gaussian distribution and add the widths in quadrature, for a total standard deviation of $0.48\,{\rm km}\,{\rm s}^{-1}\,{\rm kpc}^{-1}$. The $68$ and $95$~per~cent upper limits on the (linear) rotation curve slope are then $-1.47$ and $-0.89\,{\rm km}\,{\rm s}^{-1}\,{\rm kpc}^{-1}$, respectively. These are the values used to compute the 68~per~cent (dashed line) and 95~per~cent (solid line) confidence lower limits on the RAR slope in the lower panel of Fig.~\ref{fig-eilers}. Comparing with the upper panel, the example at higher acceleration (the dotted lines mark the same two acceleration values in both panels), where the data have a shallow slope compatible with Eq.~\ref{eq-rar} as described above, yields a constraint well below the black curve in the lower panel. The example at higher acceleration, where the data visually describe a slope steeper than is compatible with the RAR, on the other hand, correspond to a region where the slope of the fiducial relation is excluded at $>68$~per~cent confidence. We summarize the strongest constraint from this measurement, found at the lowest acceleration, corresponding to the outermost radius, with the green symbol in the lower panel: the two horizontal bars mark the $68$ and $95$~per~cent confidence lower limits, joined with an arrow. The width of the horizontal bars is representative of the $95$~per~cent confidence interval for the acceleration $g_{\rm obs}/g_\dag$ at this radius. We will use similar symbols in figures below.

We note that even incremental improvements of this measurement have the potential to turn this into a very strong limit: the exclusion curves in Fig.~\ref{fig-eilers} are \emph{extremely} sensitive to the slope of the rotation curve\footnote{E.g. the large exponent in Eq.~\ref{eq-btfr} (but note the exponent is acceleration dependent, Eq.~\ref{eq-btfr} holds exactly only in the low-acceleration limit).}. For instance, inspecting fig.~3 in \citet{2019ApJ...871..120E}, there is a (formally not significant) hint that the slope of the rotation curve may steepen to $\sim -5\,{\rm km}\,{\rm s}^{-1}\,{\rm kpc}^{-1}$ beyond $\sim 20\,{\rm kpc}$ -- were this steepening\footnote{Note that given the current constraint on the rotation speed at $20\,{\rm kpc}$, the gradient in the circular velocity cannot be steeper than about $-5\,{\rm km}\,{\rm s}^{-1}\,{\rm kpc}^{-1}$ -- any steeper corresponds to a logarithmic slope ${\rm d}\log V_{\rm circ}/{\rm d}\log r \leq -\frac{1}{2}$, which implies a declining cumulative \emph{dynamical} mass profile. For similar reasons, the slope ${\rm d}V_{\rm circ}/{\rm d}r=-1.7\,{\rm km}\,{\rm s}^{-1}\,{\rm kpc}^{-1}$ \emph{must} break to a \emph{shallower} slope before $r\sim 50\,{\rm kpc}$.} measured with confidence, the evidence for a steepening relative to the fiducial RAR at low acceleration would be at the $\sim 10\sigma$-level. Any modest improvements in the size of the confidence intervals at large radii, or the radial extent of the measurement, also have the potential to yield interesting constraints on the low-acceleration behaviour of the RAR.

\begin{figure}
  \includegraphics[width=\columnwidth]{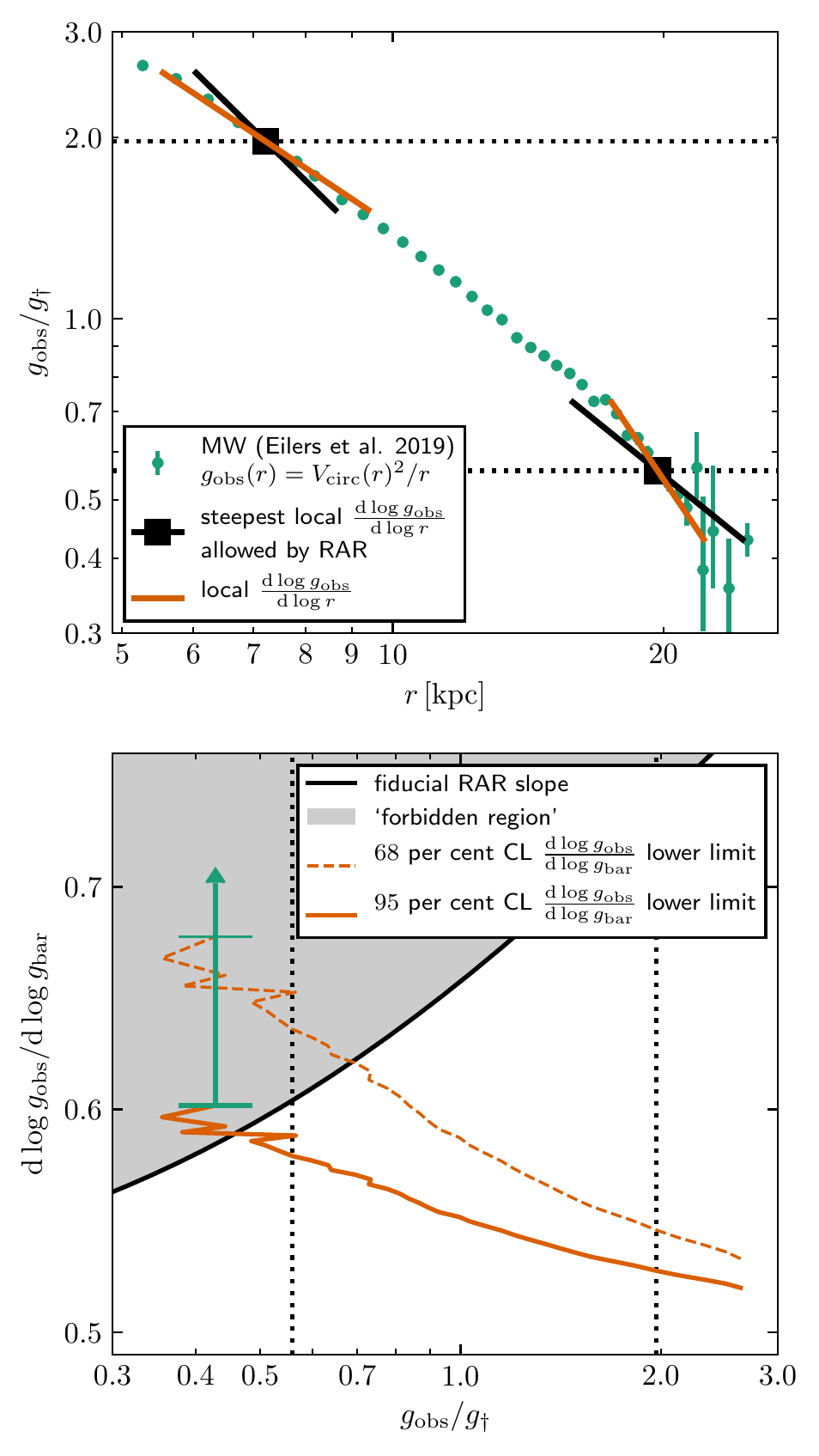}
  \caption{Constraint on the slope of the low-acceleration RAR from the Milky~Way rotation curve of \citet{2019ApJ...871..120E}. \emph{Upper panel: }The acceleration profile of the Milky~Way derived from the rotation curve as $g_{\rm obs}(r) = V_{\rm circ}(r)^2/r$. The steepest slope of the acceleration profile ${\rm d}\log g_{\rm obs}/{\rm d}\log r$ compatible with Eq.~\ref{eq-rar} is shown with a black line segment at two example accelerations, marked by the black squares and horizontal dashed lines (the vertical dashed lines in the lower panel mark the same two acceleration values). The approximate tangents to the measured acceleration profile at the same points are shown with the orange line segments; in the example at lower acceleration, the measured profile is steeper than allowed by the fiducial RAR. \emph{Lower panel: }The slope of the fiducial RAR as a function of acceleration scale is shown with the black curve (see Appendix~\ref{app-derive}), lower limits extending above this and into the grey shaded region imply a departure from Eq.~\ref{eq-rar}. The solid (dashed) lines mark the 95~per~cent (68~per~cent) confidence level lower limits on the RAR slope, as derived from the acceleration profile in the upper panel -- see Sec.~\ref{subsubsec-mwrc} for details. There is tentative evidence from the outer parts of the rotation curve (i.e. low acceleration end of lower limit curves) for a significant departure from the fiducial RAR. The most constraining lower limit (at both 68 and 95~per~cent confidence) is marked with horizontal green bars connected by an arrow.}
  \label{fig-eilers}
\end{figure}

We next turn to other probes of the low-acceleration regime.

\subsection{Weak gravitational lensing}\label{subsec-lensing}

\begin{figure*}
  \includegraphics[width=\textwidth]{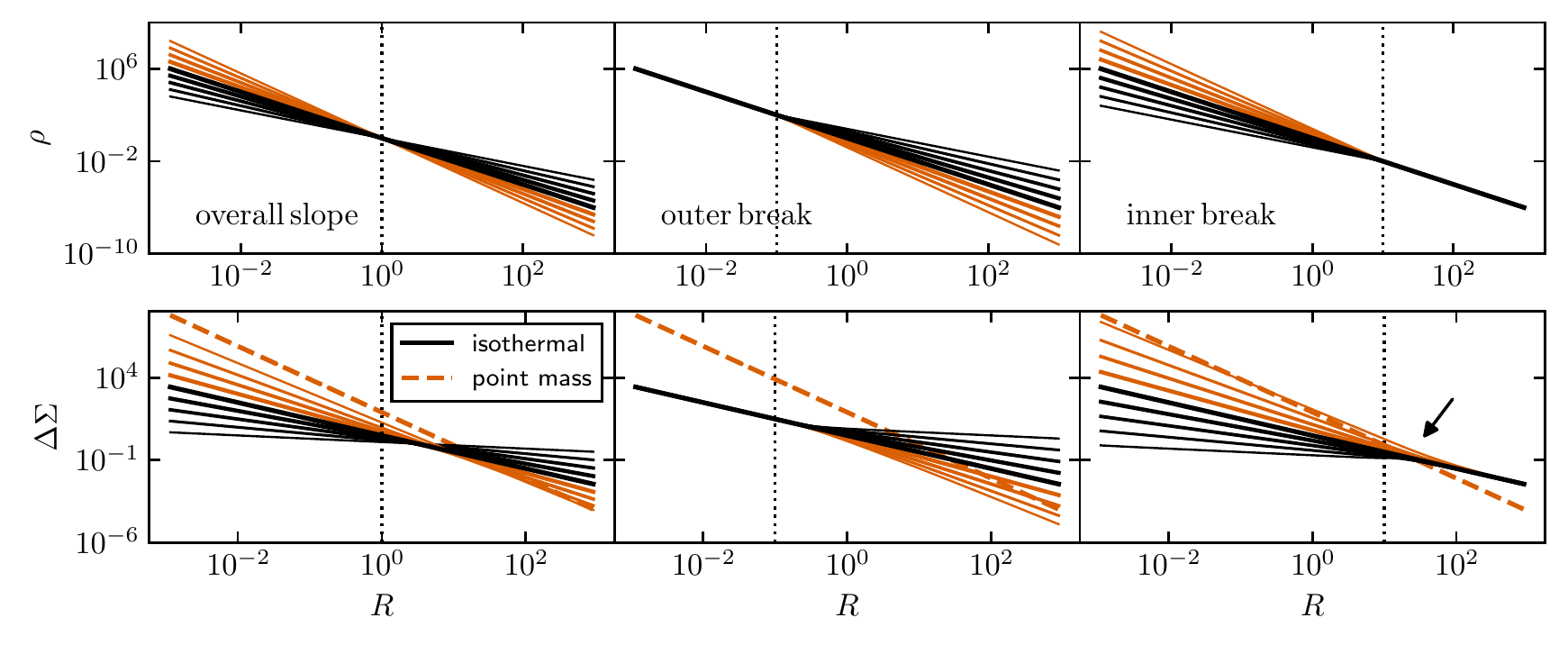}
  \caption{\emph{Left: }Power law density profiles (upper panel) and corresponding ESD profiles (lower panel). Dimensionless units and an arbitary normalization constant are assumed. Matching profiles have the same line weight and colour. The heaviest black line corresponds to an isothermal, $\rho(r)\propto r^{-2}$ profile. Steeper profiles, highlighted in orange, correspond to ESD profiles steeper than $\Delta\Sigma(R)\propto R^{-1}$, which are incompatible with the low-acceleration limit of Eq.~\ref{eq-rar}. The ESD profile for a point mass is shown with a orange dashed line. \emph{Centre: }As in left panels, but for an $\rho(r)\propto r^{-2}$ profile which breaks to a steeper (orange) or shallower (black) slope at $r>10^{-1}$. Again, orange profiles are those which conflict with the low-acceleration limit of the fiducial RAR. \emph{Right: }As in centre panels, but the slope breaks from $\rho(r)\propto r^{-2}$ at radii $r<10^{1}$. The arrow marks the region where the $\Delta\Sigma$ slope is steepened locally by the steepened $\rho(r)$ profile at interior radii, which risks a false-positive exclusion of the low-acceleration limit of Eq.~\ref{eq-rar}.}
  \label{fig-lensing}
\end{figure*}

The quantity most straightforwardly inferred from measured galaxy shapes in galaxy-galaxy weak gravitational lensing (hereafter abbreviated GGL) is the excess surface density (ESD) $\Delta\Sigma(R)=\bar{\Sigma}(R)-\Sigma(R)$, i.e. the difference between the mean surface density enclosed within radius $R$ and the local surface density at radius $R$. This is closely related to the convergence $\kappa=\Sigma/\Sigma_{\rm crit}$, where $\Sigma_{\rm crit}=\frac{4\pi G}{c^2}\frac{D_{\rm l}D_{\rm ls}}{D_{\rm s}}$, $D_{\rm l}$ and $D_{\rm s}$ are the angular diameter distance to the lens and to the source, respectively, and $D_{\rm ls}$ is the angular diameter distance between the lens and the source. While in practice it is not generally possible to uniquely infer the volume density profile $\rho(r)$ given $\Delta\Sigma(R)$, computing $\Delta\Sigma(R)$ given $\rho(r)$ is simply a matter of evaluating:
\begin{align}
  \Delta\Sigma(R)&=\frac{2}{R^2}\int_0^R\Sigma(R)R\,{\rm d}R-\Sigma(R)
\end{align}
where $\Sigma(R)=2\int_0^{\infty}\rho(\sqrt{R^2+z^2}){\rm d}z$. For the isothermal case $\rho(r)\propto r^{-2}$, the ESD is particularly simple: $\Delta\Sigma(R)\propto R^{-1}$. This is illustrated in the left panels of Fig.~\ref{fig-lensing} with the heavy black line (dimensionless units are assumed, with an arbitrary normalization constant). Another simple case is that of a point mass -- the surface density is $\Sigma(R)=0$ for all $R>0$, and the mean enclosed density falls off as $R^{-2}$, implying $\Delta\Sigma\propto R^{-2}$, illustrated with the orange dashed line in Fig.~\ref{fig-lensing}. We also show the $\Delta\Sigma$ profiles for a series of power law profiles both steeper (orange) and shallower (black) than isothermal, with greater departures from $r^{-2}$ shown with thinner lines. Under the assumption of a power-law density profile, an ESD which falls off faster than $R^{-1}$ (orange lines) implies a volume density profile which falls off too steeply to be compatible with Eq.~\ref{eq-rar} in the low-acceleration regime.

Though not all astrophysically plausible density profiles are power laws -- e.g. the NFW and most other dark matter halo models -- all may be approximated by a (possibly large) series of power law segments. In the centre and right panels of Fig.~\ref{fig-lensing}, we illustrate the $\Delta\Sigma$ profiles corresponding to two-segment power law profiles where either the inner (centre panels) or outer (right panels) density profile segment goes as $r^{-2}$. We note that only density profiles which are for some radii steeper than $r^{-2}$ -- highlighted in orange -- give rise to ESD profiles which are anywhere steeper than $R^{-1}$ -- also highlighted in orange. This suggests that GGL is a promising technique for constraining the low-acceleration slope of the RAR, especially considering that GGL measurements can extend to very large ($\gtrsim {\rm Mpc}$) radii and, consequently, low accelerations. We identify one caveat, highlighted with an arrow in the lower right panel of Fig.~\ref{fig-lensing}: a steep inner $\rho(r)$ leads to a steepening in $\Delta\Sigma(R)$ extending to somewhat larger radii. A steep $\rho(r)$ in a high-acceleration region, where this is compatible with the fiducial RAR, could therefore lead to a steep $\Delta\Sigma(R)$ in a low-acceleration region. Care must be taken to take this into account, but a sufficiently extended ESD profile should still be able to constrain the low-acceleration slope of the RAR.

We have assumed above that the gravitational lensing effect is that of general relativity. However, there are potentially appealing interpretations of the RAR which explain the correlation between $g_{\rm obs}$ and $g_{\rm bar}$ as due to a departure from the GR force law at low accelerations. We note that our method of looking for acceleration profiles which imply declining cumulative baryonic mass profiles is also applicable, for instance, in MOND: the MOND potential for a point mass leads to a lensing effect which, in the low-acceleration limit, would imply, in the context of GR, a mass profile $M_{\rm obs}(<r)\propto r$ \citep{2013PhRvL.111d1105M}, that is the `singular isothermal sphere' (SIS) profile. Adding mass to the lens at $r > 0$ must necessarily make $M_{\rm obs}(<r)$ rise more steeply, so conversely an apparent mass profile rising more gradually than the SIS implies less mass at $r > 0$, but for a point mass there is no mass at $r > 0$ to remove, which is simply a re-statement of our argument above. An analogous argument holds for the case of EG \citep{2017MNRAS.466.2547B}; whether the same argument holds in other theories of modified gravity would need to be assessed on a case-by-case basis.

Constraining the RAR using GGL requires some additional care -- particularly in ensuring that the measured region is indeed in the low-acceleration regime. The ESD profiles published in \citet[][see their fig.~3]{2017MNRAS.466.2547B} are constructed from galaxy shape measurements from the (at the time partially completed) Kilo-Degree Survey \citep[KiDS;][]{2013ExA....35...25D,2017MNRAS.465.1454H} in a region overlapping with the Galaxy and Mass Assembly \citep[GAMA;][]{2011MNRAS.413..971D,2011MNRAS.416.2640R,2015MNRAS.452.2087L} spectroscopic survey. The lens galaxies used were selected to be the most massive objects within a neighbourhood $\sim 3\,{\rm Mpc}$ in radius, reducing, but not eliminating, contamination by the acceleration field of neighbouring objects (particularly important in the context of a MOND interpretation of the RAR, see Appendix~\ref{app-efe}). Guided by Fig.~\ref{fig-lensing}, we explore below whether current GGL measurements are likely to yield interesting constraints on the low-acceleration behaviour of the RAR; a more thorough analysis leveraging the (now completed) KiDS survey data set will be presented in a companion paper (Brouwer et al., in preparation).

\begin{figure}
  \includegraphics[width=\columnwidth]{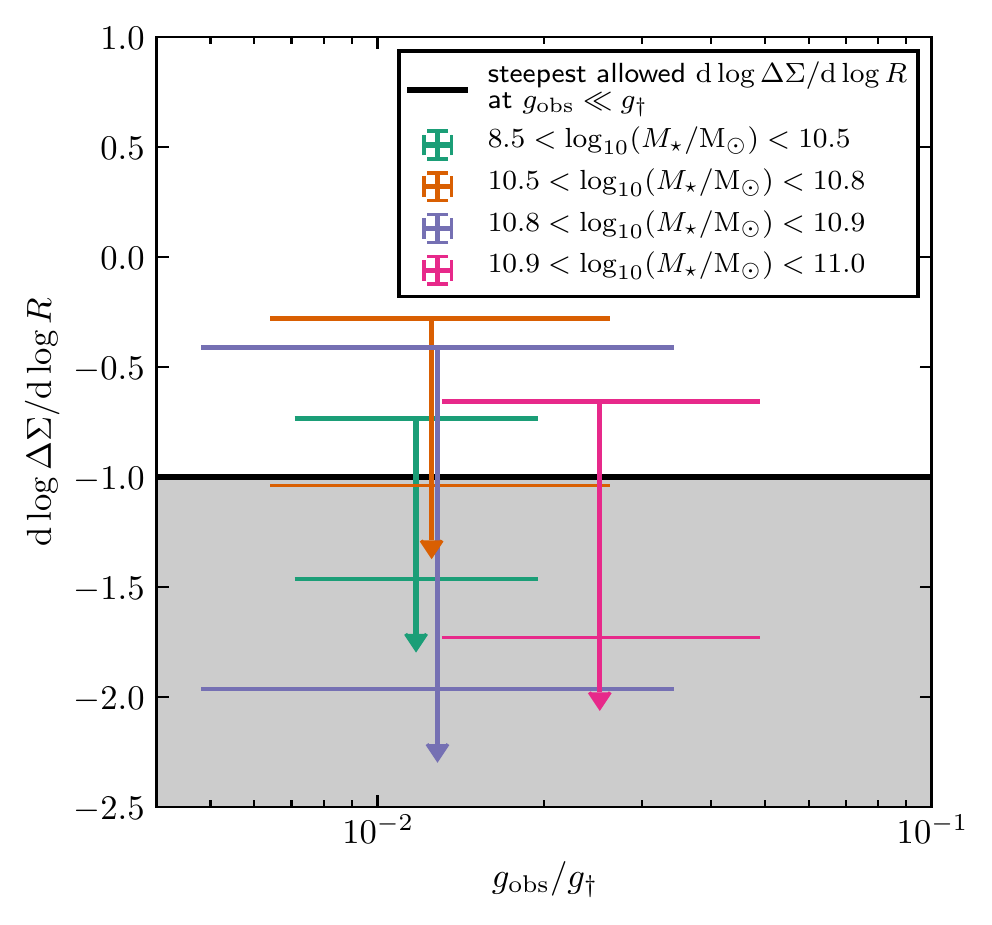}
  \caption{Constraint on the low-acceleration behaviour of the RAR from the weak lensing measurements of \citet{2017MNRAS.466.2547B}. The slope ${\rm d}\log\Delta\Sigma/{\rm d}\log R$ of their ESD profiles for lenses in $4$ stellar mass bins are estimated from the parameters of a power law constrainted by the outermost $3$ points in each profile (using more points makes no qualitative difference -- see text). We draw horizontal lines marking the $68$ and $95$~per~cent confidence upper limits; limits below ${\rm d}\log\Delta\Sigma/{\rm d}\log R=-1$ may be in tension with the low-acceleration limit of Eq.~\ref{eq-rar} (see Fig.~\ref{fig-lensing}). The acceleration scale $g_{\rm obs}/g_\dag$ is estimated using $g_{\rm obs}\approx 4G\Delta\Sigma$ (see text); the width of the symbols reflects the $95$~per~cent confidence interval estimated from the probability distributions for the power law parameters.}
  \label{fig-brouwer}
\end{figure}

In contrast to the circular velocity curve, whose slope can be directly translated to a limit on the slope of Eq.~\ref{eq-rar} (see Appendix~\ref{app-derive}), such a limit as a function of acceleration cannot be expressed in terms of the slope of the ESD profile $\Delta\Sigma(R)$ because it depends both on the local and mean enclosed surface densities. However, as detailed in the discussion surrounding Fig.~\ref{fig-lensing}, the low-acceleration limit of Eq.~\ref{eq-rar} is inconsistent with ${\rm d}\log\Delta\Sigma/{\rm d}\log R < -1$. We therefore estimate ${\rm d}\log\Delta\Sigma/{\rm d}\log R$ in the outer (i.e. lowest acceleration) parts of the ESD profiles of \citet{2017MNRAS.466.2547B} and compare with this limit in Fig.~\ref{fig-brouwer}.

We estimate ${\rm d}\log\Delta\Sigma/{\rm d}\log R$ by constraining the parameters of a power law using the three\footnote{We have repeated this analysis using from the three up to the ten outermost points, in all cases the qualitative interpretation outlined below is unchanged.} outermost measurements in each panel of fig.~3 in \citet{2017MNRAS.466.2547B}. We marginalize over the normalization parameter and plot the $68^{\rm th}$ and $95^{\rm th}$ percentiles of the distribution for the power law slope as the corresponding upper limits in Fig.~\ref{fig-brouwer}. The acceleration $g_{\rm obs}$ is in general not straightforward to estimate given only $\Delta\Sigma(R)$, however for a $\rho(r)\propto r^{-2}$ density profile, corresponding to a $\Delta\Sigma\propto R^{-1}$ ESD profile, there is a very simple relation: $g_{\rm obs}(r)=4G\Delta\Sigma(R)$, where $R=r$ (i.e. the deprojection is accounted for). The observed ESD profiles are close to $\Delta\Sigma\propto R^{-1}$, so for simplicity we assume this expression to estimate $g_{\rm obs}$, using the $95$~per~cent confidence interval for the power law fit evaluated at the radius corresponding to the $3^{\rm rd}$-outermost point in each profile.

Fig.~\ref{fig-brouwer} shows that the ESD profiles of these lens galaxies are steeper than ${\rm d}\log\Delta\Sigma/{\rm d}\log R=-1$ at $\gtrsim 68$~per~cent confidence, but not at $95$~per~cent confidence, at accelerations of $\sim 10^{-2}g_\dag$. At these accelerations the slope of Eq.~\ref{eq-rar} is very close to its low-acceleration limit (see Fig.~\ref{fig-gG}), so these measurements constitute weak evidence for a break in the RAR to a slope steeper than the fiducial value at low-acceleration. We note that we see no evidence for this measurement to be subject to the `false-positive exclusion' case illustrated in Fig.~\ref{fig-lensing}, but that ruling this out definitively requires a more in depth analysis and would be facilitated by a measurement with higher signal-to-noise ratio. Such an analysis of improved measurements will be included in Brouwer et al. (in preparation).

\subsection{Jeans modelling of resolved stars in dwarf spheroidals}\label{subsec-dsph}

At first glance, dSph galaxies seem to be ideal candidates to probe the low-acceleration behaviour of the RAR: their densities are so low that even in their centres the acceleration is, at most, comparable to $g_\dag$. However, while the low-acceleration regime is relatively easy to access via measurements of these objects, their close proximity to the Milky~Way (or Andromeda) makes the interpretation of these measurements extremely challenging. Being embedded in the tidal field of the Galaxy, it is likely that none of the dSphs discussed in this section are truly in dynamical equilibrium. Interpretation in the context of modified gravity theories such as MOND are even more challenging as even a spatially constant external acceleration field can qualitatively and anisotropically change the dynamics of an embedded system. We nevertheless derive limits on the low-acceleration slope of the RAR based on measurements of these systems, and discuss their reliability in light of these issues, below (see also Appendix~\ref{app-efe}).

We use the Jeans models for 8 dSph galaxies presented in \citet[][see their figs.~3 \& 4]{2019MNRAS.484.1401R}: Dra, Scl, For, Car, UMi, Sex, LeoI and LeoII; we omit WLM and Aqr which have qualitatively different observations and analysis, and/or very weakly constrained density profiles. \citet{2019MNRAS.484.1401R} used their Jeans modeling routine {\sc GravSphere} \citep[][see also the tests of the method by \citealp{2019arXiv191109124G}]{2017MNRAS.471.4541R,2018MNRAS.481..860R} to constrain a density profile for each dSph. The density profile is represented as a series of connected power law segments in the intervals between $0$, $\frac{1}{4}$, $\frac{1}{2}$, $1$, $2$ and $4$, times the projected half-light radius of the population of tracer stars. This profile can be described by $6$ parameters: the slope in the $5$ radial intervals and an overall normalizaton. It is clear from a visual inspection of figs.~3 \& 4 in \citet{2019MNRAS.484.1401R} that the confidence intervals for the density profiles of several of the dSphs include slopes steeper than isothermal, and in some cases that density slopes ${\rm d}\log\rho/{\rm d}\log r\geq -2$ may be excluded at high confidence.

\begin{figure}
  \includegraphics[width=\columnwidth]{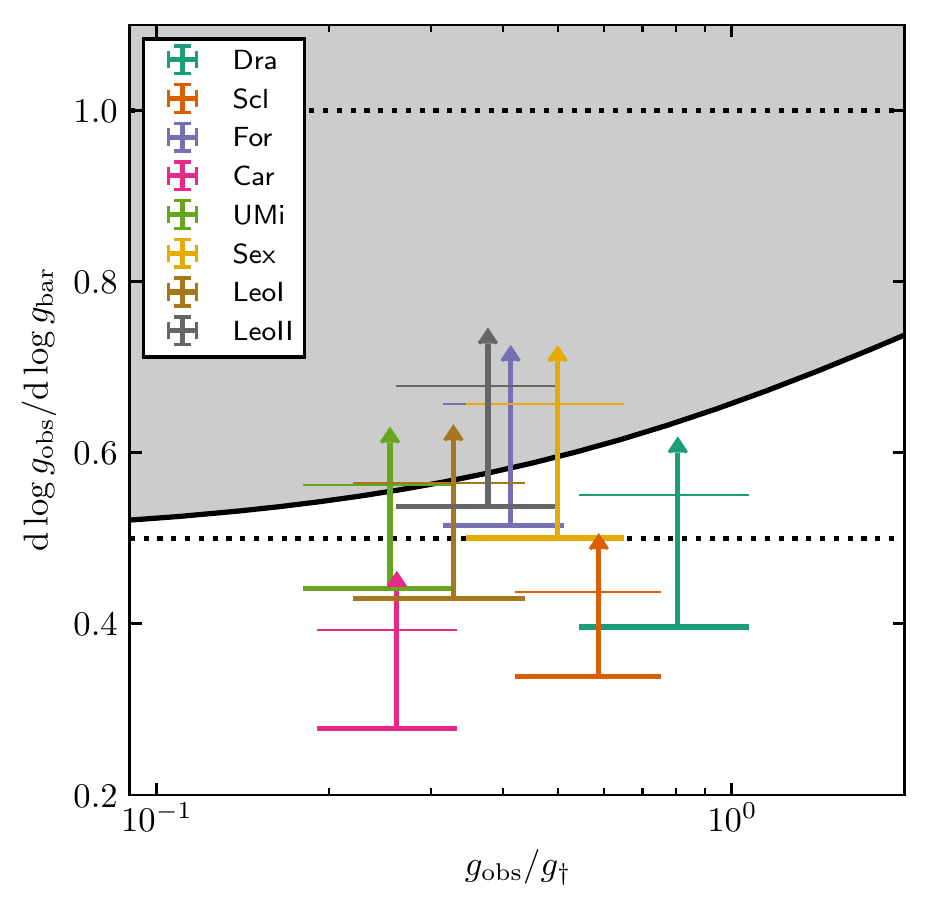}
  \caption{Constraint on the low-acceleration slope of the RAR from Jeans modelling of dSph galaxies. The slope of the fiducial relation as a function of acceleration scale is shown with the black curve (see Appendix~\ref{app-derive}); lower limits in the shaded region above this curve disfavour the fiducial RAR. Slopes less than the positions of the thick (thin) coloured horizontal bars are excluded at 95~per~cent (68~per~cent) confidence. The width of the bars is fixed by the 95~per~cent confidence interval on $g_{\rm obs}$ (obtained from the constraint on the enclosed dynamical mass in the Jeans models). The data for LeoII, Sex, For and (marginally) UMi exclude the low-acceleration slope of the fiducial RAR at $>68$~per~cent confidence; no galaxies exclude the fiducial slope at $>95$~per~cent confidence.}
  \label{fig-dsphs}
\end{figure}

The joint posterior probability distribution for the $6$ density profile parameters was estimated by \citet{2019MNRAS.484.1401R} using a Markov chain Monte Carlo estimator. We derive the marginalized constraints on the enclosed mass within each radius where the slope changes, and on the individual slopes, directly from the Markov chain samples. The measurements at the outermost radii invariably give the strongest constraint on the low-acceleration slope of the RAR, so we focus on these below. We translate the constraint on the enclosed mass $M_{\rm obs}(<r)$ into one on $g_{\rm obs}=GM_{\rm obs}(<r)/r^2$, and that on the slope of the density profile into a lower limit on the slope of the radial acceleration relation (see Appendix~\ref{app-derive}). The resulting exclusion limits on the low-acceleration RAR slope at $68$ (thin horizontal bars) and $95$~per~cent confidence (thick horizontal bars) as a function of acceleration scale are illustrated in Fig.~\ref{fig-dsphs}. There is some weak evidence ($\gtrsim 68$~per~cent confidence) from LeoII, Sex\footnote{The radial coverage of the stellar spectroscopy in Sextans extends only to $\sim 80$~per~cent of the half-light radius, whereas for the other dSphs discussed here the coverage extends at least to 2 half-light radii.}, For, UMi and LeoI for a break from the fiducial RAR toward a steeper low-acceleration slope.

Improving these constraints seems realistic: it should be mostly a matter of shrinking the confidence intervals on the density profiles of the dSphs. One straightforward (but tedious and expensive in terms of telescope time) way to achieve this is to obtain more spectra of stars in each system, but this may not even be necessary: the focus of the \citet{2019MNRAS.484.1401R} study was the inner slope of the density profile. It is conceivable that some simple adjustments to the modelling process, e.g. how the data are binned, may lead to improved constraints on the outer density slopes.

An important caveat to the above discussion is that Jeans modelling assumes a spherical system which is in dynamical equilibrium. If either of these assumptions is significantly violated, the models may not accurately reflect the actual mass profile of the system under consideration. The tests of {\sc GravSphere} by \citet{2019arXiv191109124G} suggest that the confidence intervals in Fig.~\ref{fig-dsphs} (and Fig.~\ref{fig-dsphs-efe}) are likely reliable despite the mild asphericity and departures from equilibrium in realistic dSph galaxies. However, \citet{2010ApJ...722..248M} show that the deviation of dSphs from the baryonic Tully-Fisher relation \citep[BTFR;][the RAR can be thought of as the spatially resolved analogue of the BTFR]{2000ApJ...533L..99M} anti-correlates with an estimate of how close each system is to dynamical equilibrium.

Note that in MOND, the fact that the Milky~Way dSphs are embedded in the acceleration field of the Galaxy implies a significant correction to the expected relation between $g_{\rm obs}$ and $g_{\rm bar}$, i.e. the RAR \citep{2000ApJ...541..556B}. This does not change the limits derived for the RAR slope (symbols in Fig.~\ref{fig-dsphs}), but does change the expected slope of the RAR at a given acceleration (line in Fig.~\ref{fig-dsphs}). We elaborate on this external field effect (EFE) in Appendix~\ref{app-efe}. Other modifications to gravity which predict the RAR may have qualitatively similar, but quantitatively different, corrections which would need to be accounted for on an individual basis.

\subsection{Outer Milky~Way}\label{subsec-mw}

In Sec.~\ref{subsubsec-mwrc} above, we discussed the constraint on the low-acceleration RAR slope from the Milky~Way rotation curve as determined from the rotation speeds of stars in the disc. However, other populations of dynamical tracers extend further into the outer Milky~Way and may also yield interesting constraints on the low-acceleration behaviour of the RAR. There is an extensive literature aiming broadly to constrain the mass profile of the Galaxy \citep[see][for a recent review and data compilation; see also \citealp{2020arXiv200411688S}]{2019arXiv191202599W}. We do not attempt an exhaustive consideration of every published measurement, but instead consider a few examples likely to be more constraining in the context of the argument of Sec.~\ref{sec-arg}. We note that many studies \citep[e.g.][]{2014ApJ...794...59K,2019ApJ...873..118W,2018RAA....18..113Z,2018ApJ...862...52S,2019A&A...621A..56P} assume a form of the gravitational potential (or, equivalently, density profile, circular velocity curve, etc.), usually motivated by the NFW halo profile, which excludes a $g_{\rm obs}\propto g_{\rm bar}^{\frac{1}{2}}$ low-acceleration slope \emph{ab initio}. Other studies adopt priors on the power law slope of the potential which already exclude the fiducial RAR \citep[e.g.][]{2019ApJ...875..159E,2019MNRAS.484.2832V}, or calibrate on or otherwise tie their analysis to the predictions of galaxy formation simulations assuming the $\Lambda{\rm CDM}$ cosmology \citep[e.g.][]{2008ApJ...684.1143X,2018ApJ...857...78P,2019arXiv191202086L,2019arXiv191104557C,2019MNRAS.484.5453C}, implicitly assuming that the density profile has a form similar to that found in such simulations. While these are well-motivated priors, we focus below instead on studies which allow for the possibility of radial profiles compatible with the low-acceleration limit of Eq.~\ref{eq-rar}.

We begin by considering fig.~2 in the review of \citet{2019arXiv191202599W}, which collates measurements of the mass enclosed within various radii in the outer Milky~Way from nearly $40$ individual publications. All are subject to various assumptions and possible systematic biases (and some exclude the fiducial RAR by construction, as discussed above), but the wide variety of methodologies guarantees that some, at least, are independent. The overall picture that emerges is one where the mass profile rises with a slope shallower than unity at radii corresponding to accelerations $g_{\rm obs}/g_\dag\lesssim10^{-2}$, suggesting that the RAR for the Milky~Way must bend down from the fiducial shape. We estimate constraints on the low-acceleration slope of the RAR based on a small selection of these studies below.

\begin{figure}
  \includegraphics[width=\columnwidth]{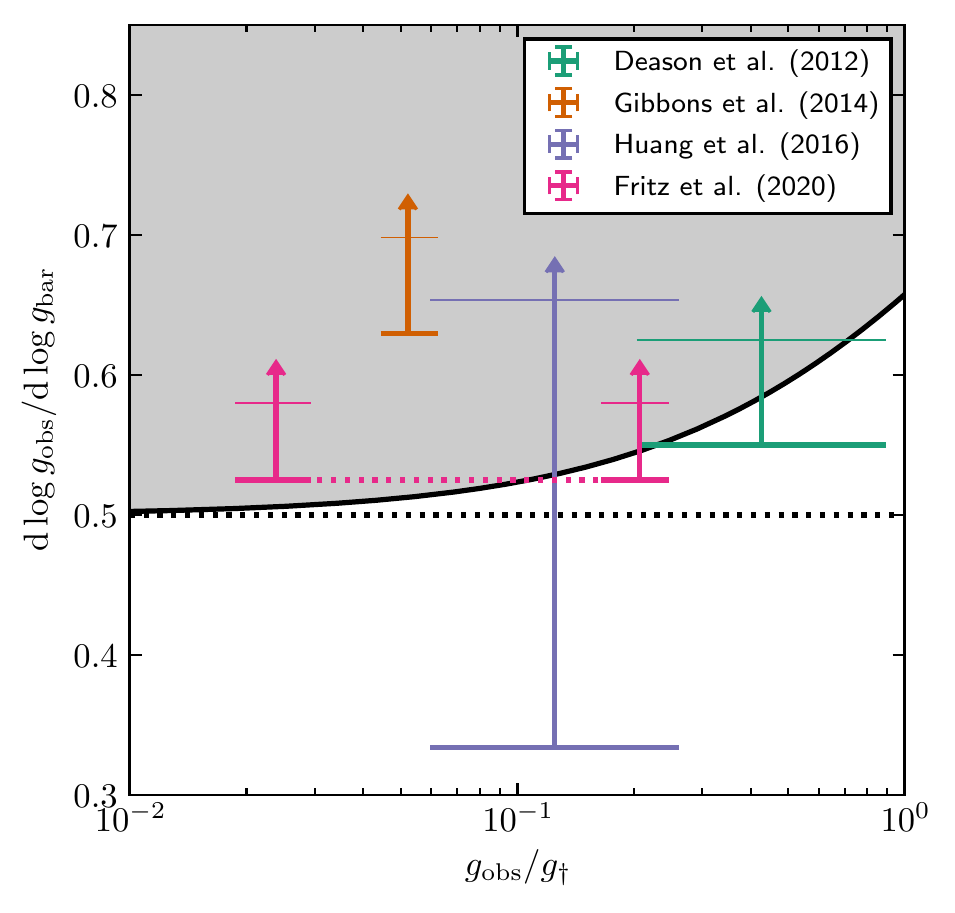}
  \caption{Constraint on the low-acceleration slope of the RAR from dynamical analyses of the outer Milky~Way. The slope of the fiducial relation as a function of acceleration scale is shown with the black curve (see Appendix~\ref{app-derive}). Lower limits in the shaded region above this curve disfavour the fiducial RAR. Symbols correspond to the constraints derived from measurements published in the works cited in the legend. Compared to Figs.~\ref{fig-eilers} \& \ref{fig-dsphs}, the limits shown here are much cruder approximations; see Sec.~\ref{subsec-mw} for details.}
  \label{fig-omw}
\end{figure}

\citet{2012MNRAS.424L..44D} used a sample of blue horizontal branch (BHB) stars in the Milky~Way stellar halo observed in the course of the Sloan Digital Sky Survey (SDSS) to constrain the parameters of a model of the distribution function for the stars. They fixed the parameters describing the density of tracers (stars) and fit only for the velocity anisotropy of tracers, and the slope and normalization of the gravitational potential. Their modelling preferred a slope\footnote{$\gamma_{\rm Deason}=0$ corresponds to a density profile $\propto r^{-2}$, while $\gamma_{\rm Deason}=1$ is the Keplerian potential.} ${\rm d}\log\Phi/{\rm d}\log r = -\gamma_{\rm Deason}$ for the gravitational potential $\Phi$ of $\gamma_{\rm Deason} \gtrsim 0.10$ $(0.25)$ at $95$~($68$)~per~cent confidence (estimated from their fig.~2) in the radial interval $16<r/{\rm kpc}<48$. The constraint is slightly weaker -- but still favours $\gamma_{\rm Deason}\sim0.35$ --  if the tracer density is assumed to be spherical rather than flattened, a possibility that the authors considered less likely. We plot the corresponding approximate $95$ and $68$~per~cent confidence lower limits on the RAR slope in Fig.~\ref{fig-omw} with the thick and thin horizontal bars, respectively; the width of the bars is set by the accelerations at $16$ and $48\,{\rm kpc}$ assuming the authors' preferred normalization of the gravitational potential ($\Phi_0=4\times10^5\,{\rm km}\,{\rm s}^{-1}$ in their notation). This measurement weakly disfavours the low-acceleration slope of the fiducial RAR.

\citet{2014ApJ...785...63B} used multiple outer Milky~Way kinematic tracers (stellar halo K-giants, BHB stars, and a heterogenous mix of globular clusters, field horizontal branch stars and dSph satellites). They treated each population separately and used a Jeans analysis to constrain the circular velocity curve in the radial interval spanned by each tracer population. They found a gentle decline from around the edge of the stellar disc ($\sim 25\,{\rm kpc}$) out to at least $\sim120\,{\rm kpc}$ (see e.g. their fig.~5), corresponding to an acceleration scale where the slope of the fiducial RAR is expected to be very close to its asymptotic value of $0.5$. As explained above, the constraint on the RAR slope is extremely sensitive to the rotation curve slope, so even the gentle decline is likely enough to argue against the fiducial RAR, although a flat slope beyond $\sim 75\,{\rm kpc}$ would seem difficult to exclude conclusively from these measurements, given the estimated errors.

In \citet{2014MNRAS.437..116B}, the authors pointed out that the rate of precession of an orbit depends on the slope of the potential; there is no precession in a Keplerian potential, while an orbit in a flat potential (corresponding to a density profile $\propto r^{-2}$) precesses by $120^\circ$ per radial period. They showed that the shape of the Sagittarius stellar stream favours a density slope steeper than $\propto r^{-2}$. \citet{2014MNRAS.445.3788G} extend the same analysis. They assumed a circular velocity curve which is flat in the inner part and has a power law slope of ${\rm d}\log V_{\rm circ}/{\rm d}\log r=-\alpha_{\rm Gibbons}/2$ in the outer part, with the transition occurring at a characteristic radius $r_{s,{\rm Gibbons}}$ as described in their eq.~4. The corresponding constraint on the RAR slope is:
\begin{align}
  \frac{{\rm d}\log g_{\rm obs}}{{\rm d}\log g_{\rm bar}}\geq \frac{\alpha_{\rm Gibbons}}{1 + \left(\frac{r_{s,{\rm Gibbons}}}{r}\right)^2}.\label{eq-gib}
\end{align}
Based on their fig.~12, we estimate $\alpha_{\rm Gibbons}=0.55\pm0.13$, $r_{s,{\rm Gibbons}}=13.9\pm6.5\,{\rm kpc}$ and $v_{0,{\rm Gibbons}}=221\pm15\,{\rm km}\,{\rm s}^{-1}$. Their measurement extends to radii of $\sim 80-100\,{\rm kpc}$, so we use this interval and the median values for $(\alpha, r_s, v_0)_{\rm Gibbons}$ to fix the acceleration scale (i.e. the width of the symbol in Fig.~\ref{fig-omw}). We estimate a conservative $>95$~($>68$)~per~cent confidence lower limit on the RAR slope by evaluating Eq.~\ref{eq-gib} at $r=90\,{\rm kpc}$ assuming values offset by 2 (1) standard deviations, for $r_{s,{\rm Gibbons}}$ upward from the median, and for $\alpha_{\rm Gibbons}$, downward -- i.e. the offsets which give the weakest constraints\footnote{These offsets run counter the clear correlation between $r_{s,{\rm Gibbons}}$ and $\alpha_{\rm Gibbons}$ \citep[][fig.~12]{2014MNRAS.445.3788G}, but accounting for the correlation would make the limit stronger, so this decision is again conservative.}. This measurement represents a very strong exclusion of the low-acceleration slope of the fiducial RAR. Even at $r\sim 20\,{\rm kpc}$, where the constraint is weaker because of the higher acceleration, and because the constraint on the circular velocity curve slope is shallower, the fiducial RAR slope is still rejected at $\geq 95$~per~cent confidence. This is intuitively easy to understand: if Eq.~\ref{eq-rar} holds, then the precession angle of the Saggitarius stream should be $120^\circ$ per radial period, a possibility which is straightforward to exclude \citep{2014MNRAS.437..116B}. However, if the precession angle per period is modified by an effect not captured in the model of \citet{2014MNRAS.445.3788G}, then this strong limit does not hold -- for instance, the predicted precession angle differs, in general, in modified gravity theories. In the particular case of MOND, for instance, the expected precession angle for the Sagittarius stream orbiting the Galactic disc (without dark matter) is nearly identical to that expected in Newtonian gravity for a Milky~Way with a mildly oblate dark matter halo, making the stream a poor discriminant between these two scenarios \citep{2005MNRAS.361..971R}.

\citet{2016MNRAS.463.2623H} measured a circular velocity for the Milky~Way combining observations of H\,{\sc i} gas, primary red clump giant and halo K-giant stars. They derived weaker constraints than \citet[][see Sec.~\ref{subsubsec-mwrc}]{2019ApJ...871..120E} at radii $\lesssim 20\,{\rm kpc}$, but the K-giant sample they used allowed them to extend their measurement out to $\sim 100\,{\rm kpc}$, so we focus here on the portion of the measurement at large radii. We estimate an upper limit on ${\rm d}\log V_{\rm circ}/{\rm d}\log r$ in the interval $40 < r / {\rm kpc} < 100$ (avoiding the putative `ring' features, see e.g. their fig.~12) as follows. We constrain the parameters of a cubic polynomial using the data in this interval assuming that the uncertainties quoted in their table~3 are Gaussian, and under the constraint that $-0.5 \leq {\rm d}\log V_{\rm circ}/{\rm d}\log r \leq 0.5$. This prior interval is limited at the lower end by the slope for a Keplerian potential and at the upper end by the plausible hypothesis that the circular velocity curve should not be sharply rising at these large radii. From the posterior distribution for the parameters of the polynomial, we derive the $68$ and $95$~per~cent lower limits on the slope as a function of radius, then take the maximum (i.e. most constraining) value along the two curves and plot the corresponding limits on the RAR slope in Fig.~\ref{fig-omw}, with the acceleration scale fixed by the interval spanned by the circular velocity curve in the radial range considered. There is clearly some evidence for a declining rotation curve -- indeed this is easily visible by eye in their fig.~12 -- but the uncertainties are large enough that it is impossible to exclude a near-flat slope with high confidence.

\citet{2020arXiv200102651F} provided constraints on the mass within $64$ and $273\,{\rm kpc}$ of the Galactic centre, and on the power-law slope of the mass profile, $\frac{{\rm d}\log M}{{\rm d}\log r}= 1-\alpha_{\rm Fritz}$, between these radii. They used the \citet{2010MNRAS.406..264W} mass estimator applied to a sample of Milky~Way satellites with proper motion measurements. They found that this estimator is biased when applied to numerical simulations of dark matter haloes from the ELVIS suite but derive a correction for this, thus implicitly assuming that an analogous bias applies when the estimator is applied to the Milky~Way data. We adopt their estimate $\alpha_{\rm Fritz}=0.27^{+0.12}_{-0.11}$ ($68$~per~cent confidence interval), and roughly estimate the $68$ and $95$~per~cent lower limits on this value as $0.27-0.11$ and $0.27-2(0.11)$, respectively. We plot the constraint on the RAR slope:
\begin{align}
  \frac{{\rm d}\log g_{\rm obs}}{{\rm d}\log g_{\rm bar}}\geq \frac{1}{2}\left(1+\alpha_{\rm Fritz}\right)\label{eq-fritz}
\end{align}
in Fig.~\ref{fig-omw} at accelerations derived from the enclosed mass\footnote{We use `case 1' from their table~1, which is the largest mass estimate and therefore the most conservative choice in the context of constraining the RAR slope.} within $64$ and $273\,{\rm kpc}$ with two symbols connected by a dotted line, where using the inner radius corresponds to the more conservative assumption. We note that the mass profile slope is in practice simply derived from the two enclosed mass estimates under the assumption that the mass profile is a power law. If the profile is assumed instead to have more freedom in shape, then the most that can be said is that it should have the average slope at some radius in the interval considered. Altogether, this measurement is suggestive of a steepening at low acceleration relative to the fiducial RAR, but depending on the interpretation perhaps only at marginal significance, and in addition dependent on validity of the assumed bias correction.

Taken altogether, the limits shown in Fig.~\ref{fig-omw} -- which we stress we have estimated rather crudely -- suggest that analyses of various dynamical tracers in the outer Milky~Way are likely either able to or on the brink of yield strong constraints on the low-acceleration behaviour of the RAR.

\section{Summary and Conclusions}\label{sec-conc}
\label{SecConc}

\citet{2016PhRvL.117t1101M} established the RAR as a tight empirical scaling relation for late-type galaxies, at least out to the lowest accelerations probed by their H\,{\sc i} gas discs. Given such an empirical relation, it is natural to explore the limits of its applicability. That the relation might depart from the extrapolation of the fiducial form of Eq.~\ref{eq-rar} was nearly immediately recognized: dSph galaxies apparently have $g_{\rm obs}$ higher than would be expected for their $g_{\rm bar}$ \citep{2017ApJ...836..152L}, but the interpretation of this fact is debated \citep[e.g. the external field effect in MOND, see Appendix~\ref{app-efe} and][]{2018MNRAS.476.3816F}. Though the measurements are more difficult than for late types, early type galaxies seem to lie on the relation \citep{2017ApJ...836..152L}, while galaxy clusters seem to lie well above the relation \citep{2020MNRAS.492.5865C}. Interestingly, $g_{\dag}$ may vary from galaxy-to-galaxy \citep{2018NatAs...2..668R,2020MNRAS.492.5865C,2020arXiv200308845Z}. The significance of the small scatter of the relation has been debated \citep[e.g.][]{2019ApJ...882....6S}, and the scatter is now known to depend on radius within each galaxy, beyond what would be expected from the radial dependence of observational uncertainties \citep{2019arXiv191109116S,2019PhRvX...9c1020R}.

In this work we have suggested a new means to search for departures from the fiducial RAR (Eq.~\ref{eq-rar}), without measuring the baryonic mass profile and independent of all uncertainties associated thereto. At low accelerations, if Eq.~\ref{eq-rar} holds and the system in question is close enough to dynamical equilibrium and spherical symmetry, then the (inversion of the) relation predicts that there are otherwise plausible acceleration profiles $g_{\rm obs}(r)$ which are unphysical on the grounds that they imply a declining cumulative baryonic mass profile (Sec.~\ref{sec-arg}). This argument can place a lower limit on the slope of the RAR, i.e. it is sensitive to breaks to steeper slopes at the low-acceleration end. The method could also be adapted to derive a limit on $g_{\dag}$ instead of the RAR slope, if this parameter is allowed to vary rather than being fixed as we have assumed throughout this work.

The physical significance of the RAR is open to interpretation. It can be seen as a tracer of the connection between a galaxy and its dark halo, arising from the intricate interplay between the masses and structural parameters of both. In modified gravity theories such as MOND or EG, the relation is instead a fundamental prediction. The low-acceleration behaviour of the RAR may be a strong discriminant between these two scenarios: in the dark matter picture, a large enough sphere must enclose the cosmic average ratio of baryons and dark matter $f_{\rm b}=\Omega_{\rm b}/\Omega_{\rm m}$ -- the RAR should then bend back down to join the line $g_{\rm obs} = f^{-1}_{\rm b}g_{\rm bar}$ (Fig.~\ref{fig-eagle-rar}) -- whereas in MOND $g_{\rm obs}\propto g_{\rm bar}^{\frac{1}{2}}$ should continue to arbitrarily low accelerations (modulo e.g. the external field effect, see Appendix~\ref{app-efe}); EG makes a similar prediction. The generic prediction for dark halo profiles has a density slope steeper than $r^{-2}$ at large radii \citep{1996ApJ...462..563N,1997ApJ...490..493N} -- a prediction which, provided the acceleration scale is low enough and baryons scarce enough, is in direct conflict with the fiducial RAR. Looking for the RAR to break to a steeper slope can therefore be viewed as qualitatively equivalent to searching for the `edge' of dark matter haloes.

Our main conclusions are summarized as follows:
\begin{itemize}
\item None of the analyses from the literature which we have considered provides definitive, model-independent evidence for a break to a RAR slope steeper than the fiducial $g_{\rm obs}\propto \sqrt{g_{\rm bar}}$ at low accelerations. However, several measurements are suggestive of such a break at $\sim 68$~per cent ($1\sigma$) confidence.
\item Rotation curves from the literature provide no compelling examples of declining rotation curves in the low-acceleration regime -- which would imply a break from the fiducial RAR -- with the possible exception of that of the Milky~Way.
\item Galaxy-galaxy weak lensing is a promising means to apply our method, and we will pursue this further in a companion paper (Brouwer et al., in prep).
\item Recent analysis of the internal dynamics of dSph satellites of the Milky~Way is on the cusp of placing strong limits on the low-acceleration RAR, but improving the constraints may be challenging\footnote{Interestingly, the dSph constraints seem to argue very strongly against a MOND interpretation: combining the demands that (i) dSphs lie on the RAR and (ii) that the low-acceleration RAR have a slope of $\frac{1}{2}$, in conjunction with accounting for the EFE, seem totally incompatible (Appendix~\ref{app-efe}), unless the dSphs are significantly out of dynamical equilibrium \citep{2010ApJ...722..248M}.}.
\item We derived constraints on the RAR slope from the results of several studies of the dynamics of objects in the outer Milky~Way -- while these are approximate, they suggest that the outskirts of the Galaxy are a promising place to look for the means to place strong constraints on the RAR slope, possibly even by leveraging existing data sets.
\end{itemize}

\section*{Acknowledgements}\label{sec-acknowledgements}

J.~Read for providing the Markov chains from \citet{2019MNRAS.484.1401R} used in Sec.~\ref{subsec-dsph}, and J.~Read and A.~Genina for guidance on their use. A.~Deason, J.~Read, J.~Pe\~narrubia, M.~Cautun, G.~Eadie, L.~Posti, C.~Frenk, A.~Fattahi, A.~Ben\'{i}tez-Llambay, T.~Theuns, A.~Genina, E.~Valentijn for useful feedback. \url{https://math.stackexchange.com} user Paul Enta for identifying an identity for integration of hypergeometric functions useful in Sec.~\ref{subsec-lensing}. KAO acknowledges support by the European Research Council (ERC) through Advanced Investigator grant to C.~S.~Frenk, DMIDAS (GA 786910). ADL acknowledges financial supported from the Australian Research Council (project number FT160100250). This research has made use of NASA's Astrophysics Data System.

\bibliography{paper}

\appendix
\section{The acceleration-dependent slope of the RAR}
\label{app-derive}

\begin{figure}
  \includegraphics[width=\columnwidth]{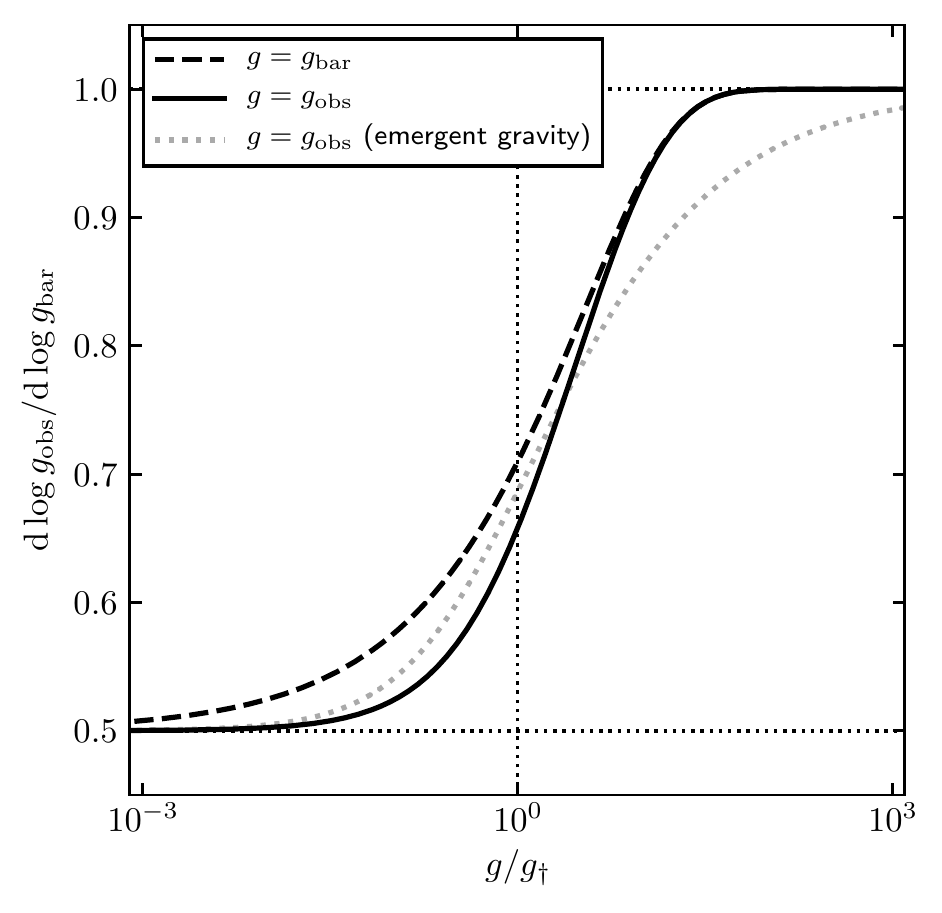}
  \caption{The logarithmic slope of the fiducial RAR (Eq.~\ref{eq-rar}) as a function of both $g_{\rm bar}$ (dashed line) and $g_{\rm obs}$ (solid line). The prediction for a the acceleration profile sourced by a point mass in emergent gravity, which has nearly the same shape, is shown with a grey dotted line.}
  \label{fig-gG}  
\end{figure}

The logarithmic slope of the RAR as expressed in Eq.~\ref{eq-rar} is:
\begin{align}
  \frac{{\rm d}\log g_{\rm obs}}{{\rm d}\log g_{\rm bar}} = 1 - \frac{1}{2}\sqrt{\frac{g_{\rm bar}}{g_\dag}}\frac{e^{-\sqrt{g_{\rm bar}/g_\dag}}}{1 - e^{-\sqrt{g_{\rm bar}/g_\dag}}},\label{eq-logslope}
\end{align}
which we plot as a function of both $g_{\rm bar}$ and $g_{\rm obs}$ in Fig.~\ref{fig-gG}. The slope does not converge to its low-acceleration limit especially quickly as a function of $g_{\rm bar}$: at $g_{\rm bar}/g_\dag=0.1$, the slope is still $\sim0.57$. It does converge somewhat more quickly as a function of $g_{\rm obs}$, since $g_{\rm obs}\geq g_{\rm bar}$.

In order to determine the limit where a given dynamical measurement implies an unphysical cumulative baryonic mass profile (assuming that Eq.~\ref{eq-rar} holds), the acceleration scale of the measurement needs to be taken into account as follows. If $M_{\rm obs}\propto r^\beta$, then the condition for a physical baryon mass profile is $\beta\geq-2\left(\frac{{\rm d}\log g_{\rm obs}}{{\rm d}\log g_{\rm bar}}-1\right)$. The equivalent constraint in terms of density, if $\rho\propto r^\gamma$, is then $\gamma\geq-\left(2\frac{{\rm d}\log g_{\rm obs}}{{\rm d}\log g_{\rm bar}}+1\right)$, or in terms of circular velocity, if $V_{\rm circ}\propto r^\zeta$, then $\zeta\geq\frac{1}{2}-\frac{{\rm d}\log g_{\rm obs}}{{\rm d}\log g_{\rm bar}}$. Note that $\alpha$, $\beta$, $\gamma$ and $\zeta$ refer to the local (at a given radius, corresponding to a given acceleration) power law slope -- in general all of these slopes are a function of radius.

Therefore, given a measurement of $M_{\rm obs}\propto r^\beta$, $\rho\propto r^\gamma$, or $V_{\rm circ}\propto r^\zeta$, the constraint on the RAR slope $g_{\rm obs}\propto g_{\rm bar}^\alpha$ is $\alpha\geq1-\frac{\beta}{2}$, $\alpha\geq-\frac{1}{2}(1+\gamma)$, or $\alpha\geq\frac{1}{2}-\zeta$, respectively. Of course this applies at the acceleration scale where the measurement is made, so measurements made at lower accelerations, where the fiducial RAR has a shallower slope, are more constraining.

The constraint in terms of the ESD profile $\Delta\Sigma(R)$, as measured from e.g. GGL, is ${\rm d}\log\Delta\Sigma/{\rm d}\log R\geq -1$ in the low-acceleration limit; the limit as a function of acceleration scale is not straightforward to express since the correspondence between $\Delta\Sigma$ and $g_{\rm obs}$ must be numerically evaluated, and in practice usually requires additional assumptions about the underlying mass profile.

We note that the acceleration-dependent slope of the acceleration profile predicted by emergent gravity (EG) for a point mass differs only slightly\footnote{It actually has exactly the same functional form as for MOND with the `simple' interpolation function as described in \citet{2012LRR....15...10F}.} from the form of Eq.~\ref{eq-logslope}. Following eq.~19 in \citet{2017MNRAS.466.2547B}, it is:
\begin{align}
  \frac{{\rm d}\log g_{\rm obs}}{{\rm d}\log g_{\rm bar}} = \frac{\sqrt{g_{\rm bar}g_\ddag}+\frac{1}{2}g_\ddag}{\sqrt{g_{\rm bar}g_\ddag}+g_\ddag}.
\end{align}
$g_\ddag$ is a constant with dimensions of acceleration. In EG, its value is predicted to be $cH_0/6\approx 1.13\times 10^{-10}\,{\rm m}\,{\rm s}^{-2}\approx 1.06 g_\dag$. We show this function with a grey dotted line in Fig.~\ref{fig-gG}. The small differences, particularly at $g_{\rm obs}\lesssim g_\dag$, imply that the constraints derived in Sec.~\ref{sec-arg} are also valid for EG, at least in the limited case where a galaxy is modelled as an isolated point particle.

\section{The MOND external field effect}
\label{app-efe}

\begin{figure}
  \includegraphics[width=\columnwidth]{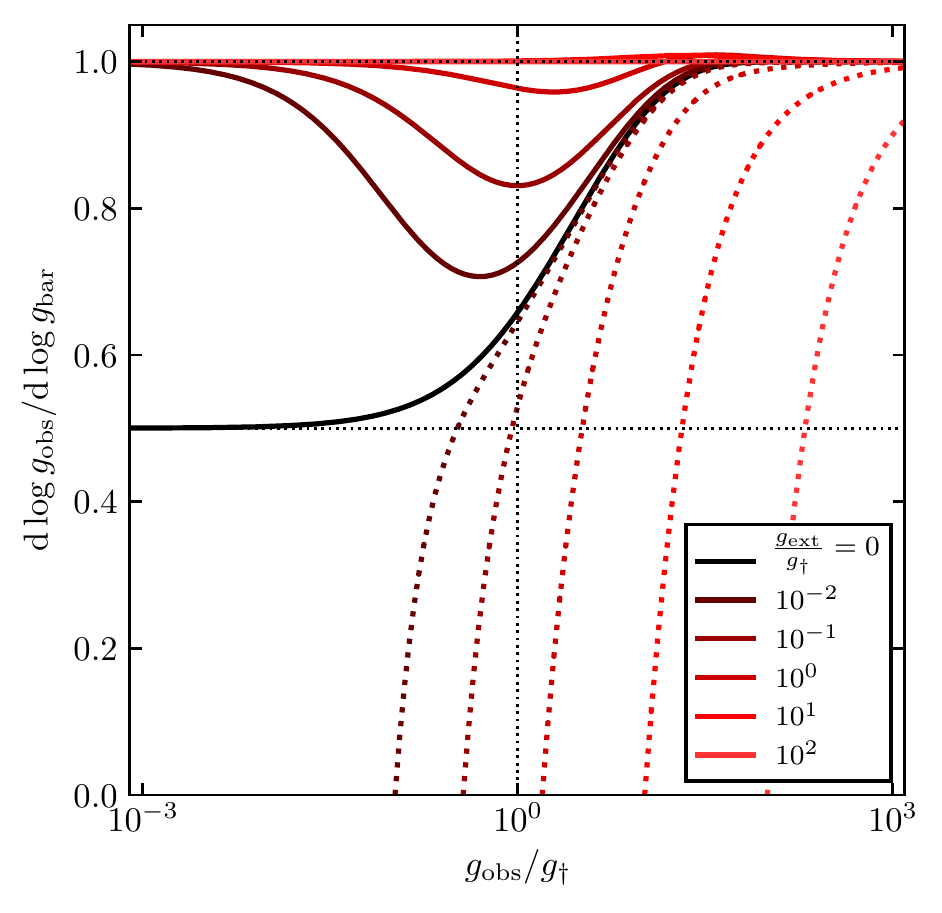}
  \caption{The slope of the radial acceleration relation as a function of $g_{\rm obs}$ in the presence of a constant external field, assuming the MOND EFE approximation given in Eq.~\ref{eq-efe}. The black solid curve from Fig.~\ref{fig-gG} is reproduced here; lighter line colours correspond to stronger external fields. The dotted curves give the RAR slopes if the `EFE-inspired' correction of \citet{2017ApJ...836..152L} is used instead (see Appendix~\ref{app-efe}).}
  \label{fig-efe}
\end{figure}

In MOND theories, the internal dynamics of a system can be significantly affected by an external field if the acceleration due to the external field $g_{\rm ext}$ is in the regime $g_{\rm bar}\lesssim g_{\rm ext}\lesssim g_\dag$. The `external field effect' (EFE) can be approximated\footnote{In addition to the fact that this is an approximate expression, as detailed in \citet{2012LRR....15...10F}, we neglect here the vector nature of the accelerations: more accurately, the EFE will have an anisotropic effect on the internal dynamics of a system. For the Milky~Way dSphs discussed later in this section, this is not wholly unreasonable: $g_{\rm obs}$ is derived from the velocity dispersion along the line of sight, and the sight line from Earth to each dSph lies nearly exactly along the direction of the external acceleration vector. However, the full effect of an anisotropic distortion of the effective acceleration field within the dSphs is unlikely to be so simple.} as given in \citet[][eq.~60]{2012LRR....15...10F}:
\begin{align}
  g_{\rm obs} = \frac{g_{\rm bar} + g_{\rm ext}}{1 - e^{-\sqrt{(g_{\rm bar}+g_{\rm ext})/g_\dag}}} - \frac{g_{\rm ext}}{1 - e^{-\sqrt{g_{\rm ext}/g_\dag}}}.\label{eq-efe}
\end{align}
This changes the slope of the RAR predicted by MOND, and so the range of allowed $g_{\rm obs}$ profiles at a given acceleration scale. The bounding curves, analogous to the solid line in Fig.~\ref{fig-gG}, for a range of external field strengths are shown with the solid curves in Fig.~\ref{fig-efe}.

\begin{figure}
  \includegraphics[width=\columnwidth]{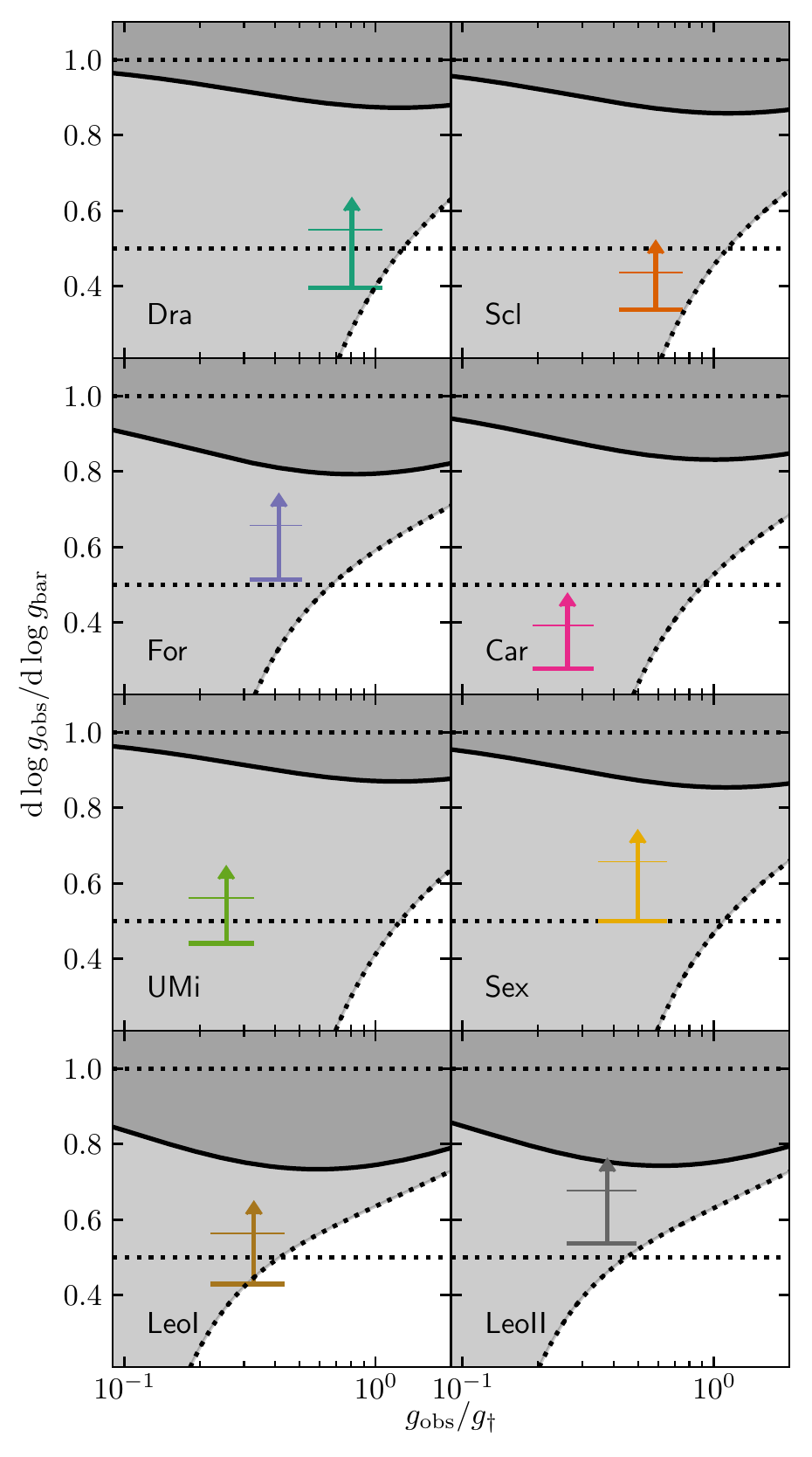}
  \caption{Analogous to Fig.~\ref{fig-dsphs}, but with the line demarking the limit of allowed RAR slopes corrected for the MOND EFE (solid black curves, see Appendix~\ref{app-efe} for details) or corrected using the `EFE-inspired' correction of \citet[][dotted curves]{2017ApJ...836..152L}.}
  \label{fig-dsphs-efe}
\end{figure}

The dSph galaxies discussed in Sec.~\ref{subsec-dsph} are relatively close to the Milky~Way and so are subject to significant EFE corrections in MOND. In Fig.~\ref{fig-dsphs-efe} we reproduce the limits from Fig.~\ref{fig-dsphs}, but with the boundary curves derived from Eq.~\ref{eq-efe} for an external field $g_{\rm ext}=GM_{\rm MW}/D_{\rm MW}^2$ where $M_{\rm MW}$ and $D_{\rm MW}$ are the (baryonic) mass of and distance to the Milky~Way, assuming the values given in \citet[][table~4]{2017ApJ...836..152L}. In all cases the constraints become extremely weak. We stress, however, that the EFE applies only in MOND: if the RAR arises for another reason, e.g. as a result of correlations between galaxy and DM halo structural parameters \citep[e.g.][]{2000ApJ...534..146V,2002ApJ...569L..19K,2016MNRAS.455..476S,2017ApJ...835L..17K,2017MNRAS.471.1841N,2017PhRvL.118p1103L,2019MNRAS.485.1886D,2019ApJ...882...46W}, in a universe with Newtonian gravity, then the correct limits are those illustrated in Fig.~\ref{fig-dsphs}.

Readers familiar with the discussion of the EFE in \citet{2017ApJ...836..152L} may remark that the weakening of the limits after correcting for the EFE seems peculiar -- in that discussion, the EFE correction resulted in an effective increase of $g_{\rm bar}$, which should make the limits in Fig.~\ref{fig-dsphs-efe} \emph{stronger} than those from Fig.~\ref{fig-dsphs}. This is because \citet{2017ApJ...836..152L} used what they term an `EFE-inspired' correction, which amounts to simply neglecting the second term on the right side of Eq.~\ref{eq-efe}. This changes the sign of the correction required to account for the external field. The slope predicted for the RAR when this alternative correction is used is shown in Fig.~\ref{fig-efe} with dotted curves, for a range of external field strengths, and the corresponding bounding curves in Fig.~\ref{fig-dsphs-efe} are shown with dotted lines. Interestingly, although this correction places the dSphs on the fiducial RAR \citep[see][fig.~11]{2017ApJ...836..152L}, doing so implies unphysical baryonic mass profiles for all 8 dSphs shown in Fig.~\ref{fig-dsphs-efe} at $\geq 95$~per~cent confidence. Conversely, if the correction from Eq.~\ref{eq-efe} is used, the slope of the EFE-corrected RAR will be allowed by the data (as discussed above), but values of $g_{\rm obs}$ and $g_{\rm bar}$ for these galaxies will be such that they lie significantly above the same EFE-corrected RAR.

Finally, we note that interpreting an application of the methodology described in Sec.~\ref{subsec-lensing} will require careful consideration of the possible impact of the EFE, although unfortunately there is no detailed treatment of lensing in the EFE limit \citep{2013PhRvL.111d1105M}.

\label{lastpage}

\end{document}